\documentclass{article}

\usepackage[utf8]{inputenc}
\usepackage[T1]{fontenc}
\usepackage{hyperref}
\usepackage{url}
\usepackage{booktabs}
\usepackage{amsfonts}
\usepackage{nicefrac}
\usepackage{microtype}
\usepackage{xcolor}
\usepackage{graphicx}
\usepackage{amsmath,amssymb,amsthm}
\usepackage{placeins}
\usepackage{natbib}

\title{A Statistical Framework for Data-Driven Discovery of Differential Performance in Clinical Risk Prediction Models}

\author{Aidan Neher$^\ast$, Julian Wolfson\\[4pt]
Biostatistics and Health Data Science, University of Minnesota \\ 
2221 University Ave SE, Minneapolis, 55414, MN, USA \\ 
{neher015@umn.edu}}

\begin{document}

\maketitle

\begin{abstract}
Predictive models employing artificial intelligence (AI) and machine learning (ML) are increasingly being used for decision support in healthcare settings. These models may exhibit differential performance across population subgroups defined by race, age, sex, and other factors and cause disparate clinical impacts, leading to intensive recent study of what has been termed "model fairness". While many methods have been proposed to assess risk prediction model fairness, these techniques generally require that the end user pre-specify the groups across which fairness is to be evaluated. In real-world settings, however, important model performance disparities may arise in unknown subgroups defined by multiple intersecting characteristics. To address this problem, we propose the unfairness tree (utree), a data-driven recursive partitioning framework for identifying subgroups with differential model performance. In simulations, the utree exhibits nominal empirical type I error rates and good ability to detect, quantify, and characterize performance discrepancies defined by higher-order variable interactions. In six mortality risk models fit to the GUSTO-I acute myocardial infarction trial dataset, utrees identified subgroup-specific performance patterns, with age, sex, blood pressure, and Killip class consistently associated with differential model performance.
\end{abstract}


\section{Introduction}\label{Introduction}

Prediction models based on artificial intelligence (AI) and machine learning (ML) tools are increasingly used to support clinical decision making. Estimates for 2023 indicate that 65\% of US hospitals are using prediction model-based tools \citep{nong_current_2025}. A risk assessment instrument (RAI) is a prediction model that outputs a risk score estimating the probability of an event (e.g., mortality or disease recurrence). As these models increasingly influence treatment allocation and clinical workflows, model unfairness---wherein predictions contribute to systematically different clinical decisions or outcomes across patient groups (e.g., undertreatment of Black patients \citep{obermeyer_dissecting_2019})---has become an important concern. Model unfairness can produce disparate health outcomes even among groups with similar risk profiles \citep{mcelfresh_call_2023, obermeyer_dissecting_2019}. 

In this work, we focus on the statistical problem of detecting and characterizing differential model performance across population subgroups for a RAI, a key initial step in determining whether a RAI is unfair. Differential model performance may arise from model misspecification, data sparsity, distributional shift, or systematic sampling biases, and can be quantified through subgroup-specific differences in calibration, discrimination, or other model performance metrics \citep{wastvedt_intersectional_2023}. Whether differential model performance results in disparate clinical impacts (i.e., unfairness) depends on the intended use of the model, the affected populations, the magnitude of performance differences across groups, and other broader ethical and clinical considerations. In the remainder of this paper, to avoid overly cumbersome verbiage and to align with common terminology usage in the fairness assessment methodology literature, we use the terms "fair" and "fairness" as a shorthand for "no statistical evidence of differential model performance". This colloquial use of the term "fairness" extends to the naming of our method which quantifies differential model performance but, for the reasons noted above, does not definitively determine whether or not a RAI is unfair in the broad sense. 

To assess whether differential model performance is present, existing approaches compare model performance across pre-specified demographic categories (e.g., race/ethnicity or sex), often relative to a designated reference group \citep{coston_counterfactual_2020, wastvedt_intersectional_2023, zink_fair_2020}. Such comparisons are important, as they can reveal potentially disparate impacts across groups known to experience health disparities. However, restricting evaluation to pre-defined subgroups may overlook disparities affecting less obvious groups defined by other characteristics or combinations thereof (e.g., individuals with a previous cancer diagnosis or females under age 18). Moreover, widely studied definitions of model fairness---including equalized odds, sufficiency, and independence---generally cannot be satisfied simultaneously (Table \ref{tab:FairnessMetrics}, \citet{mishler_fairness_2021}). As a result, a model that appears fair under one criterion or within one grouping (e.g., equalized odds across race/ethnicity relative to a reference group) may nevertheless exhibit disparities across other strata or metrics. These challenges motivate methods for data-driven discovery of subgroups with differential model performance.

We propose the unfairness tree (u-tree), a recursive partitioning framework for identifying subgroups with heterogeneous predictive discrepancies. The utree uses conditional inference trees to explain conditional (“local”) discrepancies between risk predictions and outcomes as a function of covariates \citep{hothorn_unbiased_2006}. Splitting decisions are guided by discrepancy statistics adapted from \citet{wastvedt_intersectional_2023}, which quantify the extent of differential model performance across multiple potentially intersecting subgroups. Unlike approaches that target parameter instability or require pre-specified subgroups, the proposed framework is designed to identify data-driven subgroups exhibiting differential model performance with respect to a user-defined discrepancy measure. Section \ref{ProblemDefinition} formalizes the problem of differential RAI performance and Section \ref{Methods} introduces utree, a greedy algorithm that partitions the covariate space using differential performance metrics. The development is done in the context of model calibration and we also show how to extend to common model performance metrics (Section \ref{other_metrics}). We demonstrate the approach using simulated data in Section \ref{simulation_study} and the GUSTO-I trial -- an acute myocardial infarction study -- in Section \ref{gusto_analysis}, and conclude with recommendations for the evaluation of RAIs in Section \ref{Discussion}. 

\section{Problem Definition}\label{ProblemDefinition}

Let $Y = (Y_1, \dots, Y_n)$ denote binary outcomes, where $Y_i = 1$ indicates occurrence of the event for observation $i$. For each observation, we assume access to a predicted risk $\hat{s}_i \in [0,1]$, representing the model’s estimated event probability. These risk scores are often thresholded at $\tau$ to generate binary predictions $\hat{Y}_i = \mathbb{I}(\hat{s}_i > \tau)$. Let $Z_i = (Z_{i1}, \dots, Z_{ip})^\top$ denote a vector of $p$ covariates used to assess differential model performance, and let $Z \in \mathbb{R}^{n \times p}$ denote the corresponding covariate matrix with rows $Z_i^\top$. Our goal is to evaluate whether $\hat{s}_i$ adequately approximates $\mathbb{E}[Y_i \mid Z_i]$ across subgroups defined by $Z$. The covariates used for fairness evaluation need not coincide with those used to generate the predictions; they may be a subset, superset, or entirely different set of variables, provided they are available at evaluation. To estimate stroke risk, for example, age, sex, and a history of prior stroke can be used to generate $\hat{s}$ \citep{lip_refining_2010}. However, one may wish to evaluate heterogeneity in predictive performance across covariates that encode socioeconomic status or other characteristics not explicitly included in model training.

A wide range of metrics have been proposed to measure unfairness or differential performance of RAIs, typically defined with respect to a categorical protected attribute or grouping variable $Z$. Common fairness criteria include balance in the rate of positive predictions (demographic parity), true and false positive rates (equalized odds), true positive rates (equal opportunity), positive predictive value (predictive parity), and agreement between predicted risk and outcome frequency (calibration), as summarized for binary $Z$ in Table~\ref{tab:FairnessMetrics}. These criteria are generally incompatible, and tradeoffs must be made in practice \citep{kleinberg_inherent_2016}.

We focus on \textit{model miscalibration}---discrepancies between predicted risk and observed outcomes---as a fundamental dimension of differential model performance. A model is well calibrated if predicted risks reflect true conditional outcome probabilities, i.e., $\mathbb{E}[Y \mid \hat{s}=s] = s$ for all $s \in [0,1]$. Local miscalibration arises when this relationship fails within subgroups: $\mathbb{E}[Y \mid \hat{s}=s, Z=z] \neq s$ for some $(s,z)$. We summarize this deviation using the subgroup-specific discrepancy
\begin{equation}
\delta(z)
=
\mathbb{E}[Y - \hat{s} \mid Z=z],
\end{equation}
which measures the average difference between observed outcomes and predicted risks within subgroup $z$. Under local calibration, $\delta(z)=0$ for all $z$, while positive and negative values indicate systematic underprediction and overprediction, respectively. The corresponding sample analog $\hat{\delta}(z)$ is obtained by averaging observed outcome discrepancies within the subgroup. 

In health settings, estimating $\delta(z)$ is complicated when model predictions inform treatment decisions that affect observed outcomes. Suppose a treatment $D \in \{0,1\}$ is assigned based on predicted risk $\hat{s}$. If individuals in a subgroup defined by $Z=z$ systematically receive an effective treatment ($D=1$), then the observed outcome may satisfy $\mathbb{E}[Y \mid Z=z] < \mathbb{E}[\hat{s} \mid Z=z]$. In this case, the model may appear to overestimate risk, even if it is well calibrated in the absence of treatment. This phenomenon reflects treatment-induced bias rather than true model miscalibration. To address this, we adopt a counterfactual evaluation framework from causal inference \citep{coston_counterfactual_2020}. Let $Y^0$ and $Y^1$ denote the potential outcomes under a reference condition ($D=0$) and treatment ($D=1$), respectively. We assess calibration with respect to the potential outcome $Y^0$, defining a model to be counterfactually well calibrated if $\mathbb{E}[Y^0 - \hat{s} \mid Z=z] = 0$ for all $z$. The corresponding counterfactual discrepancy is
\begin{equation}
\delta^0(z)
=
\mathbb{E}[Y^0 - \hat{s} \mid Z=z].
\end{equation}
In practice, $Y^0$ is not observed and must be estimated. We use doubly robust estimators (DR) $\hat{Y}_i^0$ that combine an outcome regression model for $\mathbb{E}[Y \mid Z, D=0]$ with a propensity score model for $\mathbb{P}(D=1 \mid Z)$. This approach integrates inverse probability weighting with outcome modeling and yields consistent estimates of $\mathbb{E}[Y^0 \mid Z]$ if either model is correctly specified \citep{coston_counterfactual_2020}. Using $\hat{Y}_i^0$ as estimates of $Y_i^0$, we can compute subgroup-specific discrepancy estimates $\hat{\delta}^0(z)$. 

\section{Methods}\label{Methods}

\subsection{Unfairness Tree Algorithm}\label{utree}

We propose the utree algorithm---a novel data-driven method that that partitions the covariate space using discrepancy estimates. We adopt the conditional-inference tree (ctree) framework \citep{hothorn_unbiased_2006}. In contrast to the more commonly applied Classification and Regression Trees (CARTs) \citep{breiman_classification_1984}, which can be biased towards the selection of continuous variables, ctrees are not biased similarly since variable and split-point selection is separated into two steps. At each node of the utree, we select the covariate and corresponding split-point that partitions the data into subgroups with maximal predictive discrepancy. Selection is guided by user defined discrepancy statistics that quantify deviation between predictions and outcomes. The recursive two-step procedure is as follows:

\begin{enumerate}

\item \textbf{Variable selection:} Compute test statistics $\mathcal{T} = \{T_1, \dots, T_p\}$ that quantify subgroup-specific discrepancy with respect to each candidate splitting covariate $Z_1, Z_2, \dots, Z_p$. Based on $\mathcal{T}$, obtain $\hat{\mathbf{p}} = (\hat{p}_1, \dots, \hat{p}_p)^\top$, a vector of empirical \textit{p}-values for the marginal null hypotheses that discrepancy does not vary across values of $Z_k$. The selected variable is the index $j$ with the smallest $\hat{p}_j$, provided it falls below multiplicity-adjusted significance level $\alpha$. If no $\hat{p}_j$ is below $\alpha$, then splitting stops at that node.

\item \textbf{Split-point selection:} For the covariate selected in Step 1, candidate split points are identified, and $\mathcal{T}$ is computed for the binary indicator variables implied by each split point (e.g., $\mathbb{1}[Z_3 < -2], \mathbb{1}[Z_3 < -1.5], \mathbb{1}[Z_3 < -1]$, etc.). The final splitting rule is determined by the split point that maximizes the magnitude of the test statistic.

\end{enumerate}

The key statistic arises from the ordered (“Lorenz”) curve, which visualizes concordance between two random variables by plotting cumulative weighted distributions. \citet{yang_double_2024} showed how the Lorenz curve can be used to assess miscalibration of a prediction model conditional on a covariate $Z_j$ by ordering pairs $(Y, \hat{s})$ according to $Z_j$ prior to constructing the curve. Under perfect calibration conditional on $Z_j$, the distribution of $Y$ along the support of $Z_j$ coincides with the distribution of predicted risk $\hat{s}$ along the same support. Formally, if calibration holds, i.e., $\delta(z)=0 \,\, \forall z$, then for any threshold $z$,
\begin{equation}
\frac{\mathbb{E}[Y \,\mathbf{1}\{Z_j \le z\}]}{\mathbb{E}[Y]} 
\;=\; 
\frac{\mathbb{E}[\hat{s}\,\mathbf{1}\{Z_j \le z\}]}{\mathbb{E}[\hat{s}]}.
\end{equation}
This suggests that a test for conditional model miscalibration can be based on the maximum difference between the empirical estimates of these expectations across values of $z$. For each candidate splitting covariate $Z_j$, this corresponds to a Kolmogorov--Smirnov test statistic comparing two weighted empirical cumulative distribution functions. Let $w^Y = Y$ and $w^{\hat{s}} = \hat{s}$ denote weights defined by the observed outcomes and predicted risks, respectively. The statistic is

\begin{equation}
\sup_z \left| 
\frac{\sum_{i=1}^n Y_i \mathbb{I}[Z_{ij} \leq z]}{\sum_{i=1}^n Y_i}
-
\frac{\sum_{i=1}^n \hat{s}_i \mathbb{I}[Z_{ij} \leq z]}{\sum_{i=1}^n \hat{s}_i}
\right|
\;\equiv\;
\sup_z \left| \widehat{F}_j(z; w^Y) - \widehat{F}_j(z; w^{\hat s}) \right|,
\label{eq:ks_stat}
\end{equation}

where $\widehat{F}_j(z; w)$ denotes the weighted empirical cumulative distribution function of the candidate splitting covariate $Z_j$ under weights $w$. Appendix~\ref{mathematical_foundations} provides additional theoretical motivation for this statistic. At each node, for a covariate $Z_j$, utree tests the null hypothesis that the weighted distributions of a candidate splitting covariate are identical under outcome and prediction weights:
\[
H_0:\ F_j(z; w^Y) = F_j(z; w^{\hat{s}})
\quad \forall z,
\]
where $F_j(z; w)$ denotes the population weighted cumulative distribution function of $Z_j$ under weights $w$. Under $H_0$, the distribution of $Z_j$ is the same when weighted by observed outcomes and predicted risks. Rejection indicates that observed and predicted risk are distributed differently across values of $Z_j$, providing evidence of subgroup-specific differential performance.

In our setting, we are interested in the (conditional) concordance between $Y^0$ and $\hat{s}$, so we replace $Y$ in \eqref{eq:ks_stat} with doubly robust estimates $\hat{Y}^0$ of potential outcomes under the reference condition. Let $w^{\hat{Y}^0} = \hat{Y}^0$ and $w^{\hat{s}} = \hat{s}$ denote weights defined by the estimated outcomes under the reference condition and predicted risks, respectively. For a candidate splitting covariate $Z_j$, the test statistic is
\begin{equation}
T_j =
\sup_z \left|
\frac{\sum_{i=1}^n \hat{Y}^0_i \,\mathbb{I}[Z_{ij} \le z]}
{\sum_{i=1}^n \hat{Y}^0_i}
-
\frac{\sum_{i=1}^n \hat{s}_i \,\mathbb{I}[Z_{ij} \le z]}
{\sum_{i=1}^n \hat{s}_i}
\right|
\;\equiv\;
\sup_z
\left|
\widehat{F}_j(z; w^{\hat{Y}^0})
-
\widehat{F}_j(z; w^{\hat{s}})
\right|.
\label{eq:utree_stat}
\end{equation}

Under calibration, the two empirical CDFs converge to the same limit; large values of $T_j$ indicate systematic divergence and hence conditional model miscalibration. 

To assess significance, we approximate the null distribution of $T_j$ by jointly permuting the pair $(\hat{Y}^0, \hat{s})$ across observations while keeping $Z$ fixed. This preserves the marginal distributions of the outcome estimates and predictions, while removing their conditional relationship with $Z$. Recomputing $T_j$ on each permuted sample yields a reference distribution against which the observed statistic is compared. The resulting empirical \textit{p}-values are used in the variable and split-point selection steps, and the algorithm proceeds recursively until a stopping criterion is met (e.g., minimum node size). These \textit{p}-values provide inference for the variable-selection procedure at the current node. In all analyses, empirical \textit{p}-values were computed using 500 permutations. Tree growth was governed by multiplicity-adjusted hypothesis testing ($\alpha=0.05$) together with a maximum tree depth of five.

\subsection{Extension to Other Performance Metrics}\label{other_metrics}

The preceding description of the utree is oriented around detecting differences in model calibration across subgroups defined by (combinations of) covariates. Here, we show how the utree framework can be formulated for other model performance metrics through the choice of weighting functions. Let $g_a, g_b : [0,1]\times[0,1]\to\mathbb{R}_{\ge 0}$. Define weights $w^{(a)} = g_a(\hat{Y}^0,\hat{s})$ and $w^{(b)} = g_b(\hat{Y}^0,\hat{s})$ where the functions are applied element-wise to the vectors of estimated outcomes under the reference condition and predicted risks. Let $\widehat{F}_j(z; w^{(a)})$ and $\widehat{F}_j(z; w^{(b)})$ denote the corresponding weighted empirical cumulative distribution functions for candidate splitting covariate $Z_j$. The absence of differential performance with respect to $Z_j$ corresponds to equality of these weighted distributions, which can be assessed using the Kolmogorov--Smirnov-type statistic
\begin{equation}
T_j(g_a,g_b)
=
\sup_z
\left|
\widehat{F}_j(z; w^{(a)})
-
\widehat{F}_j(z; w^{(b)})
\right|.
\end{equation}
By selecting $g_a$ and $g_b$, one can target different performance measures. Counterfactual calibration corresponds to $g_a=\hat{Y}^0$ and $g_b=\hat{s}$. Counterfactual true positive rate can be assessed by setting $g_a(\hat{Y}^0,\hat{s}) = \hat{Y}^0\,\mathbb{I}\{\hat{s}>\tau\}$ and $g_b(\hat{Y}^0,\hat{s}) = \hat{Y}^0$
so that $T_j(g_a,g_b)$ measures whether the distribution of predicted positives among positives varies across $Z_j$. More generally, threshold-based metrics such as precision and false positive rate can be obtained by incorporating $\mathbb{I}\{\hat{s}>\tau\}$ into the weighting functions \citep{coston_counterfactual_2020}. However, thresholding restricts attention to a subset of observations, reducing effective sample size. In all cases, large values of $T_j(g_a,g_b)$ indicate variation in the chosen performance metric across $Z_j$. In what follows, the simulation study focuses on calibration-based discrepancy to evaluate the statistical operating characteristics of the proposed framework. The GUSTO-I data analysis then illustrates the flexibility of the utree framework by applying it to calibration, true-positive rate, and precision, demonstrating how alternative discrepancy measures identify distinct patterns of subgroup-specific model performance.

\subsection{Characterizing Subgroups}\label{CharacterizingSubgroups}

When a utree splits, it produces $H$ terminal nodes (subgroups), indexed by $h = 1, \dots, H$, where $\mathcal{I}_h$ denotes the set of individuals assigned to node $h$. At this stage, the procedure has identified covariate-defined subgroups exhibiting differential model performance. We assess the predictive and structural stability of subgroup-specific discrepancy. Each terminal node yields an estimated discrepancy $\widehat{\Delta}_h$, which can be interpreted as a prediction of model miscalibration for individuals within that covariate-defined subgroup. These predictions can be evaluated in independent data by assigning each observation to its corresponding node and comparing predicted discrepancies to re-estimated discrepancies, providing a measure of out-of-sample validity. Because recursive partitioning may produce small or highly specific subgroups, discrepancy estimates in these regions can be unstable and less reproducible across data splits. To address this, we recommend assessing robustness through sensitivity analyses that examine how subgroup-level discrepancy patterns change as increasingly small terminal nodes are excluded or as alternative RAI specifications are considered. Such analyses help distinguish persistent, generalizable subgroup discrepancies from those driven by sparse data or model-specific artifacts. This sensitivity analysis approach is illustrated when we apply our model to the GUSTO-I dataset in Section \ref{gusto_analysis}. 

\section{Simulation Study}\label{simulation_study}

\subsection{Simulation Set-Up}

The purpose of these simulation studies is to assess the utree algorithm’s ability to identify and quantify model unfairness under a range of unfairness-generating scenarios. In each simulation, we generate a dataset, apply the utree, and evaluate the algorithm’s ability to detect between-group differences. Crucially, information about which specific variables create disparities in model performance is \textit{not provided a priori} to our method; only the observed data is available.

Across all simulations, covariates $Z_i=(Z_{i1},\ldots,Z_{i10})$ are generated from correlated latent normal variables. Specifically, $(Z_{i1},\ldots,Z_{i8},U_{i1},U_{i2}) \sim N(0,\Sigma)$, with exchangeable correlation $\rho \in \{0, 0.3, 0.7\}$, and binary subgroup indicators defined by $Z_{i9}=\mathbf 1\{U_{i1}>0\}$ and $Z_{i10}=\mathbf 1\{U_{i2}>0\}$ \citep{redelmeier_explaining_2020}. We define risk under the reference condition as $s_i^0 = \Pr(Y_i^0 = 1 \mid Z_i) = \mathrm{logit}^{-1}\!\left(\alpha_0 + \sum_{k=1}^{8} Z_{ik}\beta_k\right)$, with $\alpha_0=-0.5$ controlling the overall event rate and $\beta=(-1.25,\,0.07,\,-0.16,\,0.09,\,-0.095,\,0.002,\,0.093,\,0.18)^\top$ chosen to match the settings used in \citet{wastvedt_counterfactual_2024}. Following \citet{coston_counterfactual_2020}, the treated risk is $s_i^1=\Pr(Y_i^1=1\mid Z_i)=c_{\text{treat}}\,s_i^0$, with $c_{\text{treat}}=0.1$ controlling treatment effect size. Potential outcomes are simulated as $Y_i^0\sim\mathrm{Bern}(s_i^0)$ and $Y_i^1\sim\mathrm{Bern}(s_i^1)$. Treatment is simulated as $D_i \sim \mathrm{Bern}(\pi_i)$, with treatment propensity $\mathrm{logit}(\pi_i)=\mathrm{logit}(s_i^0)+1.6\,Z_{i9}$, so that propensity increases with baseline risk and is greater for the subgroup $Z_{i9}=1$. Observed outcomes are generated as $Y_i=D_iY_i^1+(1-D_i)Y_i^0$.

We distinguish between settings, which define the underlying data-generating environment, and scenarios, which define the mechanism by which unfairness is introduced into the model predictions. \textit{Settings:} Dataset sample size is varied as $N \in \{10^3, 5\times 10^3, 10^4, 5\times 10^4, 10^5\}$, and the correlation among covariates is varied as $\rho \in \{0, 0.3, 0.7\}$. The utree is applied using the counterfactual formulation, in which $\hat Y^0$ is used to estimate subgroup discrepancies. Variable and split-point selection testing is performed with a significance threshold of $\alpha=0.05$ and a Bonferroni multiple testing correction. All simulated covariates were made available as candidate splitting variables during tree construction, including both unfairness-generating variables and noise variables. \textit{Scenarios:} $s_i^0$ is perturbed to yield a simulated RAI risk score $\tilde{s}_i$. In the miscalibration mechanism, subgroup-specific miscalibration is introduced via additive logit shifts, $\mathrm{logit}(\tilde{s}_i) \leftarrow \mathrm{logit}(s_i^0) + d_1 Z_{i9} + d_2 Z_{i10} + d_{12}(Z_{i9}Z_{i10})$. Setting all $d$'s to zero yields perfect calibration and positive values induce over-estimation. In the imbalance mechanism, we impose outcome-specific shifts in the log-odds of the simulated risk score by applying outcome-conditional logit adjustments: $\mathrm{logit}(\tilde{s}_i) \leftarrow \mathrm{logit}(s_i^0)+\mathbf{1}\{Y_i^0=1\}\left(d_1^{+}Z_{i9}+d_2^{+}Z_{i10}+d_{12}^{+}(Z_{i9}Z_{i10})\right)+\mathbf{1}\{Y_i^0=0\}\left(d_1^{-}Z_{i9}+d_2^{-}Z_{i10}+d_{12}^{-}(Z_{i9}Z_{i10})\right)$. Changing $(d^{+},d^{-})$, we simulate an RAI that inflates or deflates predicted risk among individuals with $Y=1$ (positive imbalance) or among individuals with $Y=0$ (negative imbalance). Scenario severity is defined through a common magnitude parameter $m \in \{0, 0.05, 0.10, 0.20\}$ corresponding to none, small, moderate, and large unfairness, respectively. All scenarios follow the main+interaction pattern $(d_1,d_2,d_{12})=(m,0.5m,m)$. Although magnitudes are matched on the logit scale, imbalance affects only one outcome level, whereas miscalibration affects all observations. Each setting and scenario combination is simulated $B=500$ times.

To evaluate performance, we assess both variable recovery and the ability to quantify unfairness on an independent test set ($N_{\text{tst}}=10^3$). Recovery performance was summarized using the probability of recovering unfairness-generating variables and interactions across simulation replicates. To assess estimation accuracy, we compare predicted discrepancies to the true discrepancies $d_i^\star = \mathbb{E}[Y_i^0 \mid Z_i] - \tilde{s}_i$, where $\tilde{s}_i$ is the simulated risk score. Each test observation is assigned a leaf-wise estimate from the fitted tree (or the global mean if no splits occur), and performance is summarized using mean squared error (MSE) and bias across test observations. We additionally record computation time for fitting the utree, with variable and split-point selection implemented serially.

\subsection{Simulation Results}

\begin{figure}[!h]
\centering
\caption{
Probability of selecting at least one split as a function of sample size. Panels correspond to combinations of unfairness mechanism (miscalibration, negative imbalance, positive imbalance) and severity (none, small, moderate, large). Lines represent Monte Carlo averages across $B=500$ replicates. Split probability increases with both sample size and severity, while remaining controlled near the nominal level under the null.
}
\label{fig:nonnull_split_prob}

\includegraphics[width=\textwidth]{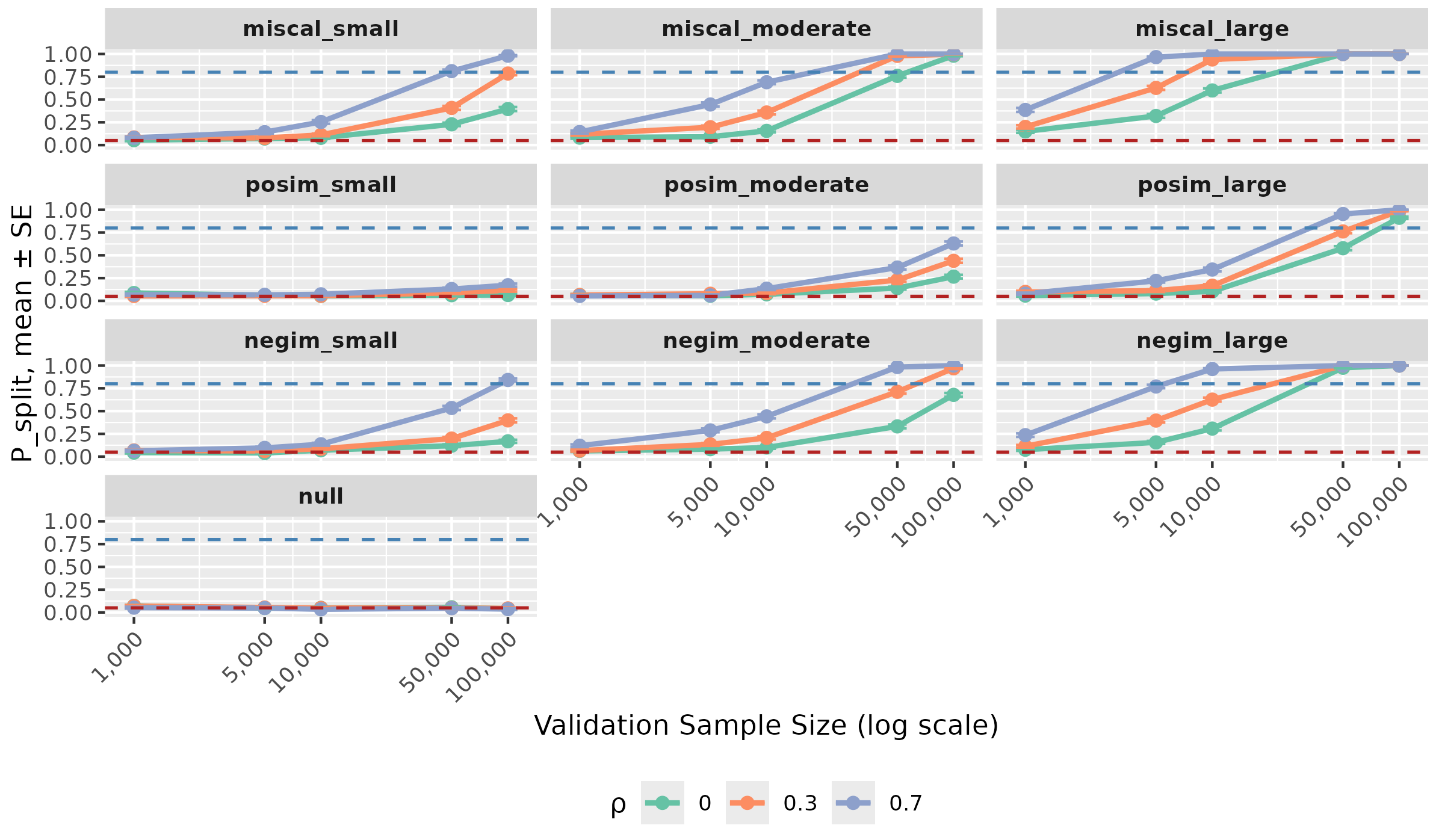}
\end{figure}

Under the null scenario, using the counterfactual framework with the calibration test statistic (Figure \ref{fig:nonnull_split_prob}), the probability of selecting a split remains close to the nominal level of 0.05 (e.g., 0.048 at $N = 10^3$, Table \ref{tab:cf_obs_comparison}), indicating empirical Type I error control. Under non-null settings, the probability of selecting a split increases with both sample size and discrepancy severity, reflecting the method’s ability to detect meaningful heterogeneity. While detection remains conservative in weaker signal settings---due to multiplicity from formal hypothesis testing and adaptive partitioning---power improves substantially with increasing signal strength, with high detection rates under strong miscalibration (e.g., 0.94 at $N = 10^3$). 

Beyond detection, the utree effectively identifies the sources of discrepancy. Variable recovery rates increase markedly with both sample size and severity (Figure \ref{fig:nonnull_var_recovery}), rising from 0.060 under small miscalibration to 0.924 under large miscalibration at $N = 10^3$, demonstrating that the method can accurately localize unfairness when signal is present. Recovery of unfairness-generating variables was generally higher than recovery of unfairness-generating interactions, reflecting the greater difficulty of recovering higher-order subgroup structure. Interaction recovery improved with increasing sample size and discrepancy severity but remained lower than variable recovery across most simulation settings (Figure~\ref{fig:nonnull_interaction_recovery}). Correlation between unfairness generating variables and noise variables through $\rho$ increased the probability of variable and interaction recovery. Recovery performance differed across discrepancy-generating mechanisms, with strongest performance observed under calibration-focused mechanisms and more modest recovery under the imbalance settings, particularly at smaller sample sizes and weaker signal strengths.

Furthermore, the utree provides accurate estimates of discrepancy magnitude. Leaf-wise estimates exhibit small bias that decreases with increasing sample size (Figure \ref{fig:nonnull_bias}), indicating approximate unbiasedness and consistency, particularly under strong miscalibration and positive imbalance. Finally, computational performance scales predictably with sample size (Figure \ref{fig:nonnull_fit_time}), with fitting remaining feasible even at $n = 10^5$ (less than 4 minutes across all settings and scenarios).

\section{Real Data Application: GUSTO-I trial}\label{gusto_analysis}

\subsection{GUSTO-I Dataset}

We leverage the GUSTO-I trial as a well-studied setting to evaluate risk assessment instruments (RAIs). The trial enrolled 41,021 patients with acute myocardial infarction (AMI) across 15 countries and 1,081 hospitals between December 27, 1990 and February 22, 1993 \citep{gusto_international_1993}. The primary outcome is 30-day mortality. Patients were randomized to one of four treatment arms; we restrict attention to those receiving streptokinase (SK) or tissue plasminogen activator (tPA), yielding a sample of $n=30{,}510$. This dataset has been widely used for developing and evaluating mortality prediction models \citep{lee_predictors_1995, steyerberg_clinical_2019}.

Using the utree framework, we evaluated six risk models developed on the GUSTO-I dataset \citep{lee_predictors_1995,steyerberg_clinical_2019,gusto_international_1993}. We divided the data into training (20\%), evaluation (60\%), and test (20\%) sets. We first fit models from the literature on the training set with 30-day mortality as the binary outcome. Four logistic regression models were specified: a bivariate additive model of age and Killip class (a prognostic factor; Age+Killip), the same model also with an interaction (Age*Killip), an additive model with 17 demographic, clinical, and treatment predictors (Table \ref{tab:utree_variable_dictionary}; Lee Simplified), and the same model with nonlinear terms for age and height (Lee Full) \citep{lee_predictors_1995}. Additionally, we fit a Generalized Additive Model (GAM) with nonlinear effects for age, height, weight, pulse, and systolic blood pressure, and a random forest to the 17 predictors.

Similar global predictive performance was observed across the six mortality prediction models, with discrimination ranging from approximately 0.78--0.82 and generally good overall calibration (Table~\ref{tab:gusto_model_performance}). We performed data-driven subgroup discovery using counterfactual utrees, with all 17 predictors considered at variable selection. To demonstrate the flexibility of the proposed framework, we repeated the subgroup discovery procedure using three discrepancy statistics: calibration discrepancy, true-positive rate (TPR), and precision. Each discrepancy statistic defines a distinct notion of differential model performance while using the same recursive partitioning algorithm. Treatment was defined as receipt of tPA ($D=1$) versus SK ($D=0$), so that $Y^0$ denotes the potential 30-day mortality outcome under SK treatment. Utrees were grown with Bonferroni-adjusted significance thresholds ($\alpha = 0.05$) and maximum tree depth of five. Variable-selection p-values were estimated using 500 permutations. Doubly robust estimates were obtained using logistic regression for the outcome and propensity models. Cross-fitting was not employed. For calibration, node-level discrepancies were summarized using the normalized relative discrepancy $100 \times \frac{\mathbb{E}(\hat{Y}_0 - \hat{s})}{\mathbb{E}(\hat{s})}$, where $\mathbb{E}(\hat{s})$ denotes the model-wide average predicted risk in the evaluation set. For TPR and precision, node-level discrepancies were summarized as relative deviations from the corresponding model-wide performance metric. The held-out test set was used to assess the stability of subgroup discrepancy findings. Because subgroup definitions were learned using the evaluation set, discrepancy estimates computed in the evaluation set should be interpreted as discovery-stage estimates. Test-set analyses and pruning sensitivity analyses were used to evaluate the reproducibility of identified subgroup discrepancy patterns in independent data.

\subsection{Data Analysis Results}

\begin{figure}[htbp]
    \centering
    \includegraphics[width=0.85\textwidth]{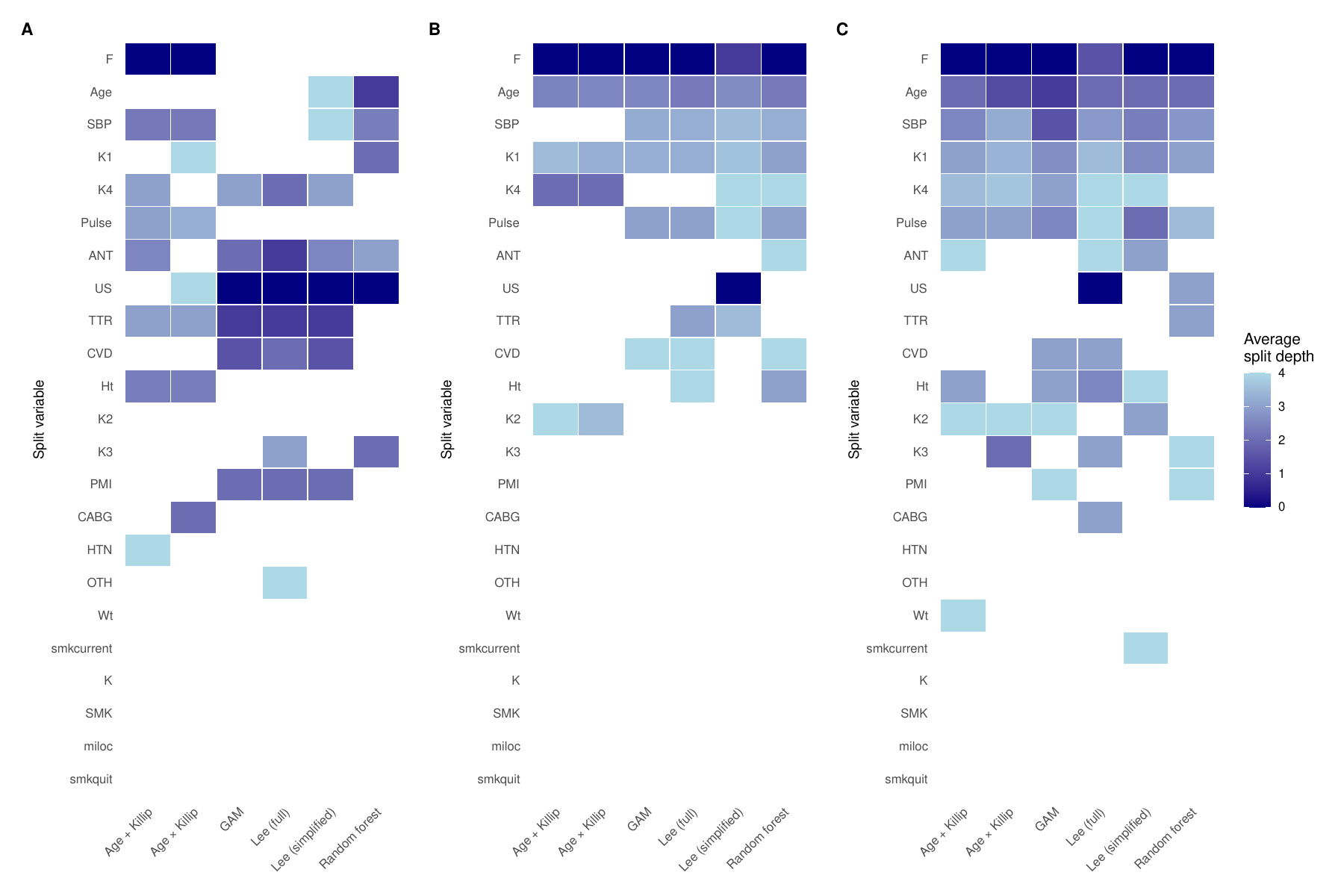}
    \caption{Split variables selected by unfairness trees across prediction models using (A) calibration discrepancy, (B) true-positive rate (TPR), and (C) precision. Rows correspond to candidate splitting variables and columns to prediction models. Cell shading indicates the average depth at which a variable was selected within a tree, with darker colors indicating earlier (more influential) splits; white cells indicate that the variable was not selected. Comparing panels illustrates how the variables selected to define subgroups differ across performance metrics. Variables selected consistently at shallow depths explain differences in model performance regardless of model choice.}
    \label{fig:split_variable_heatmap}
\end{figure}

Patterns of subgroup definition depended on the performance metric (Figure~\ref{fig:split_variable_heatmap}). Across all three performance metrics, age, female sex (F), systolic blood pressure (SBP), and Killip class (particularly K1 and K4) were consistently selected across multiple prediction models, suggesting that these characteristics are broadly associated with heterogeneous model performance. Calibration-based trees additionally and consistently selected U.S. enrollment (USe), time to treatment relief (TTR), anterior infarct location (ANT), and prior cardiovascular disease (CVD), whereas TPR-based trees relied primarily on the shared core variables. Precision-based trees likewise selected the core variables but incorporated a broader set of secondary predictors, including height (HT), prior myocardial infarction (PMI), and additional Killip indicators, although these appeared less consistently across models. Together, these findings indicate that while certain patient characteristics are consistently associated with differential model performance, the subgroup structures identified by utrees depend on the chosen performance metric.

\begin{figure}[htbp]
    \centering
    \includegraphics[width=\textwidth]{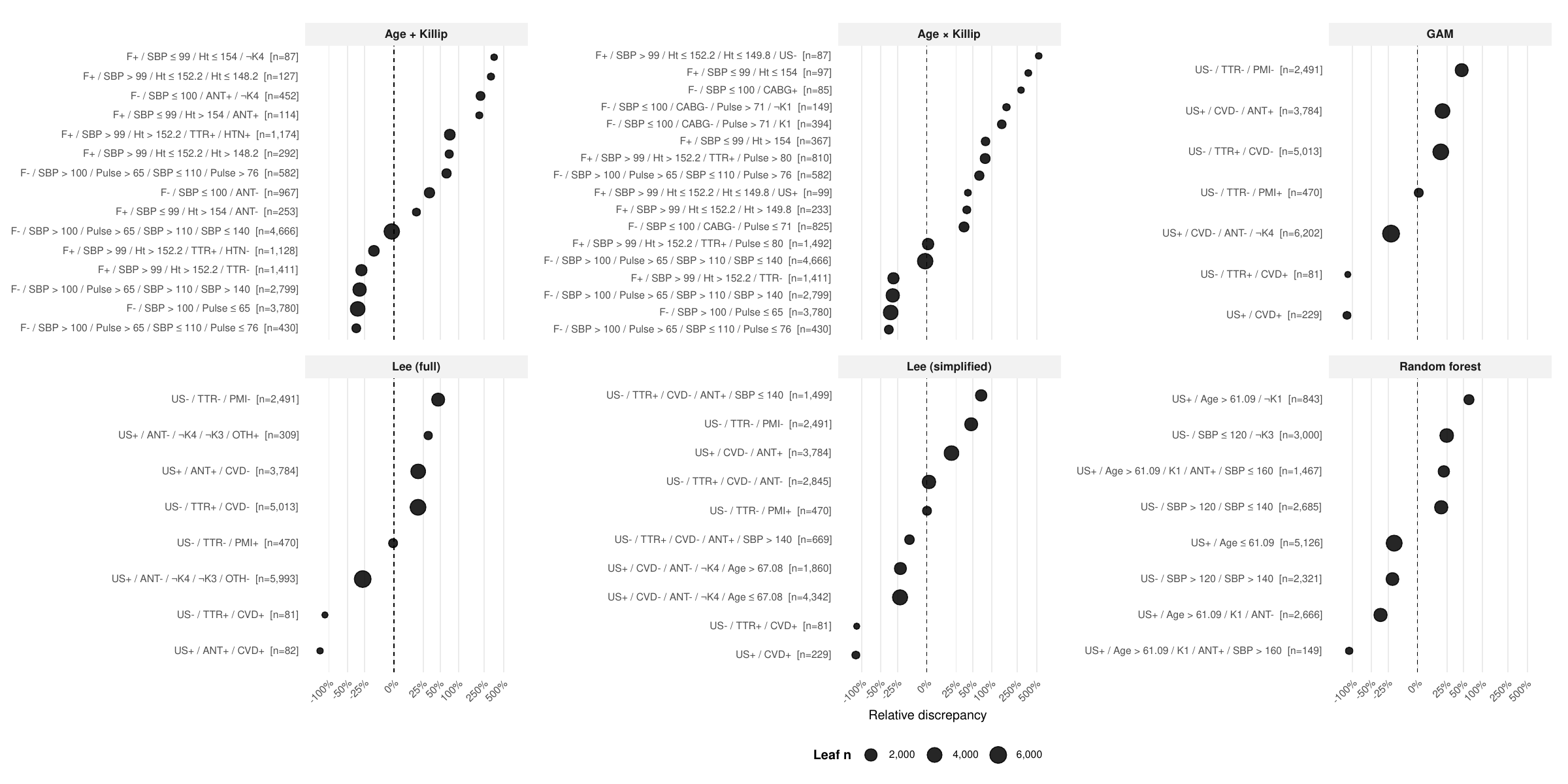}
    \caption{
    Terminal subgroups identified by calibration-based unfairness trees fit to each prediction model. Points denote terminal subgroups positioned by relative calibration discrepancy, $100\times \mathbb{E}(\hat{Y}_0-\hat{s}) / \mathbb{E}(\hat{s})$, where $\mathbb{E}(\hat{s})$ is the model-wide average predicted risk. Positive values indicate underprediction ($\hat{Y}_0>\hat{s}$), whereas negative values indicate overprediction. Point size is proportional to subgroup size. Terminal nodes with fewer than 50 observations are omitted. The dashed vertical line denotes zero discrepancy. Subgroup signatures summarize the sequence of splits defining each terminal node; ``+'' and ``$-$'' denote the presence or absence of binary characteristics.
    }
    \label{fig:utree_leaf_discrepancy}
\end{figure}

Despite similar aggregate performance characteristics, calibration-based utrees revealed substantial heterogeneity in subgroup calibration across prediction models (Figure~\ref{fig:utree_leaf_discrepancy}). Several terminal subgroups exhibited positive relative discrepancy, indicating underprediction in which observed outcome risk exceeded model-predicted risk. For example, the calibration-based utree for the Lee (full) model identified a large subgroup of non-U.S. patients with prolonged time to treatment relief and no prior cardiovascular disease (US$-$/TTR$+$/CVD$-$; $n=5{,}013$) exhibiting substantial underprediction. In contrast, a similarly large subgroup of U.S. patients without anterior or other infarct locations and without Killip class III or IV symptoms (US$+$/ANT$-$/K4$-$/K3$-$/OTH$-$; $n=5{,}993$) exhibited overprediction. These subgroup signatures summarize the sequence of recursive splits defining each terminal node, illustrating how combinations of routinely collected clinical characteristics identify heterogeneous model calibration. 

The simpler logistic models (Age + Killip and Age $\times$ Killip) tended to exhibit larger and more numerous extreme subgroup discrepancies, whereas the Lee models, GAM, and random forest generally produced more moderate relative discrepancies, although clinically meaningful subgroup deviations remained. Across models, most subgroup definitions involved interactions among three or more variables, suggesting that differential model performance frequently arises from combinations of patient characteristics rather than simple one-variable partitions. Analogous analyses based on true-positive rate (TPR) and precision identified distinct subgroup structures, exemplifying that the subgroup characteristics identified by utrees depend on the chosen performance metric (Figure~\ref{fig:utree_leaf_discrepancy_all_metrics}).

Out-of-sample validation suggested that many subgroup performance patterns were reproducible across independent data splits. Weighted correlations between evaluation- and test-set leaf estimates remained high across all three discrepancy metrics, although sensitivity to minimum terminal leaf size varied by metric and prediction model (Figure~\ref{fig:leaf_size_sensitivity}). A minimum terminal leaf size of $n\geq50$ provided a practical balance between validation stability and subgroup retention, with the greatest gains observed for calibration-based trees.

\section{Discussion}\label{Discussion}

We introduce the unfairness tree (utree), a data-driven framework for identifying, localizing, and quantifying differential performance in clinical risk models. By combining recursive partitioning, discrepancy-based test statistics, and counterfactual outcome estimation, the utree provides a principled and interpretable approach to detecting subgroup-specific miscalibration. This addresses a key limitation of existing AI fairness assessment approaches which are typically restricted to pre-defined categories and may overlook disparities in less obvious or intersecting populations.

Simulations demonstrate important properties of the utree. First, the method maintains appropriate Type I error control under the null, with split probabilities close to the nominal level (e.g., 0.048 at $N=10^3$), despite the adaptivity inherent in recursive partitioning. Second, the utree exhibits increasing power to detect differential performance as sample size and discrepancy severity increase. Third, the method reliably identifies variables driving subgroup unfairness. Fourth, the utree provides accurate estimates of discrepancy magnitude, with leaf-wise discrepancy estimates exhibiting small, diminishing bias as sample size increases. These results indicate that the utree is a statistically valid and consistent framework for detecting and characterizing subgroup-specific performance disparities.

% Inherent limitations
Limitations warrant consideration. First, subgroup discrepancies identified by the utree should be interpreted as indicators of differential model performance rather than algorithmic unfairness. Whether such discrepancies constitute unfairness depends on the intended application, affected populations, and the fairness criterion of interest. Second, the counterfactual formulation relies on standard causal identification assumptions, including positivity, consistency, no unmeasured confounding, and correct specification of either the propensity score or outcome regression model. Consequently, counterfactual subgroup discrepancies should be interpreted cautiously in observational settings where treatment assignment may depend on unmeasured factors such as clinician judgment or patient preferences, and are most compelling in randomized studies or settings where these assumptions are plausible.

% Future work
Several methodological extensions remain for future work. Although Section~\ref{other_metrics} demonstrates how the utree framework can target alternative performance measures through user-specified weighting functions, the simulation study focuses exclusively on calibration-based discrepancy. Evaluating operating characteristics, including type I error control, subgroup recovery, discrepancy estimation, and stability, for alternative discrepancy measures such as true-positive rate and precision remains an important direction for future work. Likewise, although multiplicity-adjusted permutation testing provides family-wise error rate control for variable selection at each node, this guarantee does not extend to the probability that a particular variable appears somewhere in the final tree because the tree is constructed recursively using data-dependent splits. Formal theoretical guarantees for subgroup stability under recursive partitioning also remain an open question, despite encouraging empirical validation across independent data splits. Finally, because terminal subgroups are identified through an adaptive discovery procedure, the resulting discrepancy estimates should be interpreted as exploratory summaries rather than post-selection valid estimators. Developing selective inference procedures that provide confidence intervals and hypothesis tests accounting for adaptive subgroup discovery represents another promising avenue for future research.

\bibliographystyle{plainnat}
\bibliography{references}

@misc{kleinberg_inherent_2016,
	title = {Inherent {Trade}-{Offs} in the {Fair} {Determination} of {Risk} {Scores}},
	url = {http://arxiv.org/abs/1609.05807},
	doi = {10.48550/arXiv.1609.05807},
	urldate = {2025-09-18},
	publisher = {arXiv},
	author = {Kleinberg, Jon and Mullainathan, Sendhil and Raghavan, Manish},
	month = nov,
	year = {2016},
	note = {arXiv:1609.05807 [cs]},
}

@article{lip_refining_2010,
	title = {Refining {Clinical} {Risk} {Stratification} for {Predicting} {Stroke} and {Thromboembolism} in {Atrial} {Fibrillation} {Using} a {Novel} {Risk} {Factor}-{Based} {Approach}: {The} {Euro} {Heart} {Survey} on {Atrial} {Fibrillation}},
	volume = {137},
	issn = {0012-3692},
	shorttitle = {Refining {Clinical} {Risk} {Stratification} for {Predicting} {Stroke} and {Thromboembolism} in {Atrial} {Fibrillation} {Using} a {Novel} {Risk} {Factor}-{Based} {Approach}},
	url = {https://www.sciencedirect.com/science/article/pii/S0012369210600670},
	doi = {10.1378/chest.09-1584},
	number = {2},
	urldate = {2025-09-18},
	journal = {Chest},
	author = {Lip, Gregory Y. H. and Nieuwlaat, Robby and Pisters, Ron and Lane, Deirdre A. and Crijns, Harry J. G. M.},
	month = feb,
	year = {2010},
	pages = {263--272},
}

@article{breiman_classification_1984,
	title = {Classification {And} {Regression} {Trees}},
	doi = {https://doi.org/10.1201/9781315139470},
	language = {en},
	author = {Breiman, Leo},
	year = {1984},
}

@article{gusto_international_1993,
	title = {An {International} {Randomized} {Trial} {Comparing} {Four} {Thrombolytic} {Strategies} for {Acute} {Myocardial} {Infarction}},
	volume = {329},
	issn = {0028-4793},
	url = {https://www.nejm.org/doi/full/10.1056/NEJM199309023291001},
	doi = {10.1056/NEJM199309023291001},
	number = {10},
	urldate = {2025-03-19},
	journal = {New England Journal of Medicine},
	publisher = {Massachusetts Medical Society},
	author = {GUSTO},
	month = sep,
	year = {1993},
	note = {\_eprint: https://www.nejm.org/doi/pdf/10.1056/NEJM199309023291001},
	pages = {673--682},
}

@article{hothorn_unbiased_2006,
	title = {Unbiased {Recursive} {Partitioning}: {A} {Conditional} {Inference} {Framework}},
	volume = {15},
	issn = {1061-8600},
	shorttitle = {Unbiased {Recursive} {Partitioning}},
	url = {https://doi.org/10.1198/106186006X133933},
	doi = {10.1198/106186006X133933},
	number = {3},
	urldate = {2025-04-21},
	journal = {Journal of Computational and Graphical Statistics},
	publisher = {ASA Website},
	author = {Hothorn, Torsten and , Kurt, Hornik and and Zeileis, Achim},
	month = sep,
	year = {2006},
	note = {\_eprint: https://doi.org/10.1198/106186006X133933},
	pages = {651--674},
}

@inproceedings{redelmeier_explaining_2020,
	title = {Explaining {Predictive} {Models} with {Mixed} {Features} {Using} {Shapley} {Values} and {Conditional} {Inference} {Trees}},
	isbn = {978-3-030-57321-8},
	issn = {1611-3349},
	url = {https://link.springer.com/chapter/10.1007/978-3-030-57321-8_7},
	doi = {10.1007/978-3-030-57321-8_7},
	language = {en},
	urldate = {2025-04-17},
	booktitle = {Machine {Learning} and {Knowledge} {Extraction}},
	publisher = {Springer, Cham},
	author = {Redelmeier, Annabelle and Jullum, Martin and Aas, Kjersti},
	year = {2020},
	pages = {117--137},
}

@article{yang_double_2024,
	title = {Double {Probability} {Integral} {Transform} {Residuals} for {Regression} {Models} with {Discrete} {Outcomes}},
	volume = {33},
	issn = {1061-8600, 1537-2715},
	url = {http://arxiv.org/abs/2308.15596},
	doi = {10.1080/10618600.2024.2303336},
	language = {en},
	number = {3},
	urldate = {2025-03-26},
	journal = {Journal of Computational and Graphical Statistics},
	author = {Yang, Lu},
	month = jul,
	year = {2024},
	note = {arXiv:2308.15596 [stat]},
	pages = {787--803},
}

@article{obermeyer_dissecting_2019,
	title = {Dissecting racial bias in an algorithm used to manage the health of populations},
	volume = {366},
	issn = {0036-8075, 1095-9203},
	url = {https://www.science.org/doi/10.1126/science.aax2342},
	doi = {10.1126/science.aax2342},
	language = {en},
	number = {6464},
	urldate = {2025-03-19},
	journal = {Science},
	author = {Obermeyer, Ziad and Powers, Brian and Vogeli, Christine and Mullainathan, Sendhil},
	month = oct,
	year = {2019},
	pages = {447--453},
}

@article{nong_current_2025,
	title = {Current {Use} {And} {Evaluation} {Of} {Artificial} {Intelligence} {And} {Predictive} {Models} {In} {US} {Hospitals}},
	volume = {44},
	issn = {0278-2715},
	url = {https://www.healthaffairs.org/doi/abs/10.1377/hlthaff.2024.00842},
	doi = {10.1377/hlthaff.2024.00842},
	number = {1},
	urldate = {2025-03-19},
	journal = {Health Affairs},
	publisher = {Health Affairs},
	author = {Nong, Paige and Adler-Milstein, Julia and Apathy, Nate C. and Holmgren, A. Jay and Everson, Jordan},
	month = jan,
	year = {2025},
	pages = {90--98},
}

@article{zink_fair_2020,
	title = {Fair regression for health care spending},
	volume = {76},
	issn = {1541-0420},
	url = {https://onlinelibrary.wiley.com/doi/abs/10.1111/biom.13206},
	doi = {10.1111/biom.13206},
	language = {en},
	number = {3},
	urldate = {2025-02-13},
	journal = {Biometrics},
	author = {Zink, Anna and Rose, Sherri},
	year = {2020},
	note = {\_eprint: https://onlinelibrary.wiley.com/doi/pdf/10.1111/biom.13206},
	pages = {973--982},
}

@article{mcelfresh_call_2023,
	title = {A call for better validation of opioid overdose risk algorithms},
	volume = {30},
	issn = {1527-974X},
	url = {https://doi.org/10.1093/jamia/ocad110},
	doi = {10.1093/jamia/ocad110},
	number = {10},
	urldate = {2025-02-07},
	journal = {Journal of the American Medical Informatics Association},
	author = {McElfresh, Duncan C and Chen, Lucia and Oliva, Elizabeth and Joyce, Vilija and Rose, Sherri and Tamang, Suzanne},
	month = oct,
	year = {2023},
	pages = {1741--1746},
}

@article{lee_predictors_1995,
	title = {Predictors of 30-{Day} {Mortality} in the {Era} of {Reperfusion} for {Acute} {Myocardial} {Infarction}},
	volume = {91},
	url = {https://www.ahajournals.org/doi/full/10.1161/01.cir.91.6.1659},
	doi = {10.1161/01.CIR.91.6.1659},
	number = {6},
	urldate = {2025-01-29},
	journal = {Circulation},
	publisher = {American Heart Association},
	author = {Lee, Kerry L. and Woodlief, Lynn H. and Topol, Eric J. and Weaver, W. Douglas and Betriu, Amadeo and Col, Jacques and Simoons, Maarten and Aylward, Phil and Van de Werf, Frans and Califf, Robert M.},
	month = mar,
	year = {1995},
	pages = {1659--1668},
}

@inproceedings{coston_counterfactual_2020,
	address = {New York, NY, USA},
	series = {{FAT}* '20},
	title = {Counterfactual risk assessments, evaluation, and fairness},
	isbn = {978-1-4503-6936-7},
	url = {https://dl.acm.org/doi/10.1145/3351095.3372851},
	doi = {10.1145/3351095.3372851},
	urldate = {2025-01-29},
	booktitle = {Proceedings of the 2020 {Conference} on {Fairness}, {Accountability}, and {Transparency}},
	publisher = {Association for Computing Machinery},
	author = {Coston, Amanda and Mishler, Alan and Kennedy, Edward H. and Chouldechova, Alexandra},
	month = jan,
	year = {2020},
	pages = {582--593},
}

@book{steyerberg_clinical_2019,
	address = {Cham},
	series = {Statistics for {Biology} and {Health}},
	title = {Clinical {Prediction} {Models}: {A} {Practical} {Approach} to {Development}, {Validation}, and {Updating}},
	copyright = {http://www.springer.com/tdm},
	isbn = {978-3-030-16398-3 978-3-030-16399-0},
	shorttitle = {Clinical {Prediction} {Models}},
	url = {http://link.springer.com/10.1007/978-3-030-16399-0},
	doi = {10.1007/978-3-030-16399-0},
	language = {en},
	urldate = {2024-08-14},
	publisher = {Springer International Publishing},
	author = {Steyerberg, Ewout W.},
	year = {2019},
}

@inproceedings{mishler_fairness_2021,
	address = {New York, NY, USA},
	series = {{FAccT} '21},
	title = {Fairness in {Risk} {Assessment} {Instruments}: {Post}-{Processing} to {Achieve} {Counterfactual} {Equalized} {Odds}},
	isbn = {978-1-4503-8309-7},
	shorttitle = {Fairness in {Risk} {Assessment} {Instruments}},
	url = {https://dl.acm.org/doi/10.1145/3442188.3445902},
	doi = {10.1145/3442188.3445902},
	urldate = {2024-08-02},
	booktitle = {Proceedings of the 2021 {ACM} {Conference} on {Fairness}, {Accountability}, and {Transparency}},
	publisher = {Association for Computing Machinery},
	author = {Mishler, Alan and Kennedy, Edward H. and Chouldechova, Alexandra},
	month = mar,
	year = {2021},
	pages = {386--400},
}

@article{wastvedt_intersectional_2023,
	title = {An intersectional framework for counterfactual fairness in risk prediction},
	issn = {1465-4644, 1468-4357},
	url = {http://arxiv.org/abs/2210.01194},
	doi = {10.1093/biostatistics/kxad021},
	language = {en},
	urldate = {2024-04-26},
	journal = {Biostatistics},
	author = {Wastvedt, Solvejg and Huling, Jared and Wolfson, Julian},
	month = aug,
	year = {2023},
	note = {arXiv:2210.01194 [stat]},
	pages = {kxad021},
}

@misc{wastvedt_counterfactual_2024,
	title = {Counterfactual fairness for small subgroups},
	url = {http://arxiv.org/abs/2310.19988},
	language = {en},
	urldate = {2024-04-26},
	publisher = {arXiv},
	author = {Wastvedt, Solvejg and Huling, Jared D. and Wolfson, Julian},
	month = jan,
	year = {2024},
	note = {arXiv:2310.19988 [stat]},
}

\appendix

\setcounter{figure}{0}
\setcounter{table}{0}

\renewcommand{\thefigure}{A\arabic{figure}}
\renewcommand{\thetable}{A\arabic{table}}

\section{Acknowledgments}
% For anonymous submissions, the \texttt{ack} environment is hidden automatically. It reappears in final and preprint modes.

The authors thank Solvejg Wastvedt and Jared Huling for valuable conversations and methodological insights that contributed to the development of this work. We also gratefully acknowledge the University of Minnesota Center for Learning Health System Sciences (CLHSS) for institutional support.

\section{Conflicts of Interest}

Aidan Neher is employed by Blue Cross Blue Shield of Minnesota. This work was conducted independently as part of doctoral research at the University of Minnesota. Blue Cross Blue Shield of Minnesota had no role in the study design, analysis, interpretation of the results, manuscript preparation, or decision to publish. Julian Wolfson declares no competing interests.

\section{Funding}

No funding was received specifically for this work.

\section{Background on Fairness Metrics}

Definitions in Table \ref{tab:FairnessMetrics} formalize different notions of equitable model performance across subgroups defined by a protected or clinically relevant characteristic $Z$. Because many fairness criteria are mathematically incompatible \citep{kleinberg_inherent_2016}, the choice of metric depends on the intended clinical application and the type of disparity under consideration. In this work, we focus primarily on calibration-based notions of fairness, which assess whether predicted risks correspond to outcome frequencies within subgroups.

\begin{table}[ht]
\centering
\caption{Common fairness metrics for risk assessment instruments.}
\begin{tabular}{ll}
\hline
\textbf{Metric} & \textbf{Definition} \\
\hline
Demographic parity &
$P(\hat{Y}=1 \mid Z=0) = P(\hat{Y}=1 \mid Z=1)$ \\
Equalized odds &
$P(\hat{Y}=1 \mid Y=y, Z=0) = P(\hat{Y}=1 \mid Y=y, Z=1), \; y \in \{0,1\}$ \\
Equal opportunity &
$P(\hat{Y}=1 \mid Y=1, Z=0) = P(\hat{Y}=1 \mid Y=1, Z=1)$ \\
Predictive parity &
$P(Y=1 \mid \hat{Y}=1, Z=0) = P(Y=1 \mid \hat{Y}=1, Z=1)$ \\
Calibration within groups &
$\mathbb{E}[Y \mid \hat{s}, Z=z] = \hat{s}, \;\; \forall z$ \\
\hline
\end{tabular}
\label{tab:FairnessMetrics}
\end{table}

\section{Mathematical Foundations of the Unfairness Tree Test Statistic}\label{mathematical_foundations}

\subsection{Introduction}

The unfairness tree (utree) algorithm uses a Kolmogorov-Smirnov (KS) type test to identify subgroups where a prediction model exhibits differential performance. This section provides the mathematical bridge between the conceptual motivation---detecting local miscalibration---and the specific distributional test implemented in the algorithm. 

\subsection{Null Hypotheses}

Recall the subgroup-specific calibration discrepancy function introduced in the main text:
\[
\delta_j(z)
=
\mathbb{E}[Y-\hat{s}\mid Z_j=z]
=
m_Y(z)-m_s(z),
\]
where
\[
m_Y(z)=\mathbb{E}[Y\mid Z_j=z],
\qquad
m_s(z)=\mathbb{E}[\hat{s}\mid Z_j=z].
\]

Under perfect calibration conditional on $Z_j$,
\[
\delta_j(z)=0
\qquad \forall \ z\in \mathrm{supp}(Z_j),
\]
equivalently,
\[
m_Y(z)=m_s(z)
\qquad \forall \ z.
\]

\subsection{Weighted CDF Formulation}

The unfairness tree statistic is based on comparing cumulative weighted distributions of the candidate splitting covariate $Z_j$. Define the population weighted cumulative distribution functions
\[
F_j(z;w^Y)
=
\frac{
\mathbb{E}[Y\,\mathbf{1}\{Z_j\le z\}]
}{
\mathbb{E}[Y]
},
\]
and
\[
F_j(z;w^{\hat{s}})
=
\frac{
\mathbb{E}[\hat{s}\,\mathbf{1}\{Z_j\le z\}]
}{
\mathbb{E}[\hat{s}]
}.
\]

The normalized KS null tested by the unfairness tree procedure is
\[
H_0:
F_j(z;w^Y)
=
F_j(z;w^{\hat{s}})
\qquad \forall z.
\]

Using the law of total expectation,
\[
\frac{
\int_{-\infty}^{z} m_Y(u)\, dP_{Z_j}(u)
}{
\mathbb{E}[Y]
}
=
\frac{
\int_{-\infty}^{z} m_s(u)\, dP_{Z_j}(u)
}{
\mathbb{E}[\hat{s}]
}
\qquad \forall z.
\]

Equivalently,
\[
\int_{-\infty}^{z}
\left\{
\frac{m_Y(u)}{\mathbb{E}[Y]}
-
\frac{m_s(u)}{\mathbb{E}[\hat{s}]}
\right\}
dP_{Z_j}(u)
=
0
\qquad \forall z.
\]

If this equality holds for all thresholds $z$, then the integrand must be zero almost surely with respect to $P_{Z_j}$:
\[
\frac{m_Y(z)}{\mathbb{E}[Y]}
=
\frac{m_s(z)}{\mathbb{E}[\hat{s}]}
\qquad \text{a.s.}
\]

Therefore,
\[
m_Y(z)
=
\lambda\, m_s(z)
\qquad \text{a.s.},
\]
where
\[
\lambda
=
\frac{\mathbb{E}[Y]}{\mathbb{E}[\hat{s}]}.
\]

Thus, equality of the weighted CDFs corresponds to a relative calibration condition: conditional observed risk is proportional to conditional predicted risk across the support of $Z_j$.

\subsection{Interpretation}

% Connection to subgroup discrepancy

The cumulative discrepancy associated with $Z_j$ satisfies
\[
\mathbb{E}\!\left[
(Y-\hat{s})\mathbf{1}\{Z_j\le t\}
\right]
=
\int_{-\infty}^{t}
\delta_j(u)\,
dP_{Z_j}(u).
\]

Consequently, if $\delta_j(z)$ differs systematically from zero over some region of the covariate space, the cumulative weighted distributions of observed and predicted risk diverge, resulting in a larger KS statistic. This statistic therefore targets heterogeneous subgroup-specific calibration structure.

% Relative versus absolute calibration

Furthermore, the normalized KS formulation is insensitive to uniform multiplicative calibration shifts shared across all subgroups. For example, if
\[
m_Y(z)=1.1\,m_s(z)
\qquad \forall z,
\]
then
\[
\lambda = 1.1,
\]
and the normalized KS null still holds. In this setting, the model globally underpredicts risk by 10\%, but there is no subgroup-specific differential performance with respect to $Z_j$.

Conversely, constant additive discrepancy generally violates the normalized KS null. Suppose
\[
m_Y(z)=m_s(z)+c,
\]
for some constant $c\neq 0$. Then
\[
\frac{m_Y(z)}{m_s(z)}
=
1+\frac{c}{m_s(z)},
\]
which varies with $z$ whenever $m_s(z)$ varies across the support of $Z_j$. Intuitively, a constant additive discrepancy affects low-risk groups proportionally more than high-risk groups. For example, adding $0.05$ to a subgroup with predicted risk $0.05$ doubles risk, whereas adding $0.05$ to a subgroup with predicted risk $0.50$ increases risk by only 10\%. Although the absolute calibration error is constant, the relative calibration error varies across the support of $Z_j$, inducing subgroup-specific differences detectable by the normalized KS statistic.

Consequently, the unfairness tree methodology is best interpreted as a test for heterogeneous relative miscalibration rather than a test for overall global calibration. By normalizing the weighted CDFs, the procedure removes uniform global bias from consideration and instead partitions the data only when calibration error varies systematically across candidate subgroups.

\section{Additional Simulation Results}

\begin{figure}[!h]
\centering
\caption{
Probability of recovering the true unfairness-driving variable under non-null settings as a function of sample size. Panels correspond to combinations of unfairness mechanism and severity. Lines represent Monte Carlo averages across $B=500$ replicates. Variable recovery improves with both sample size and severity.
}
\label{fig:nonnull_var_recovery}

\includegraphics[width=\textwidth]{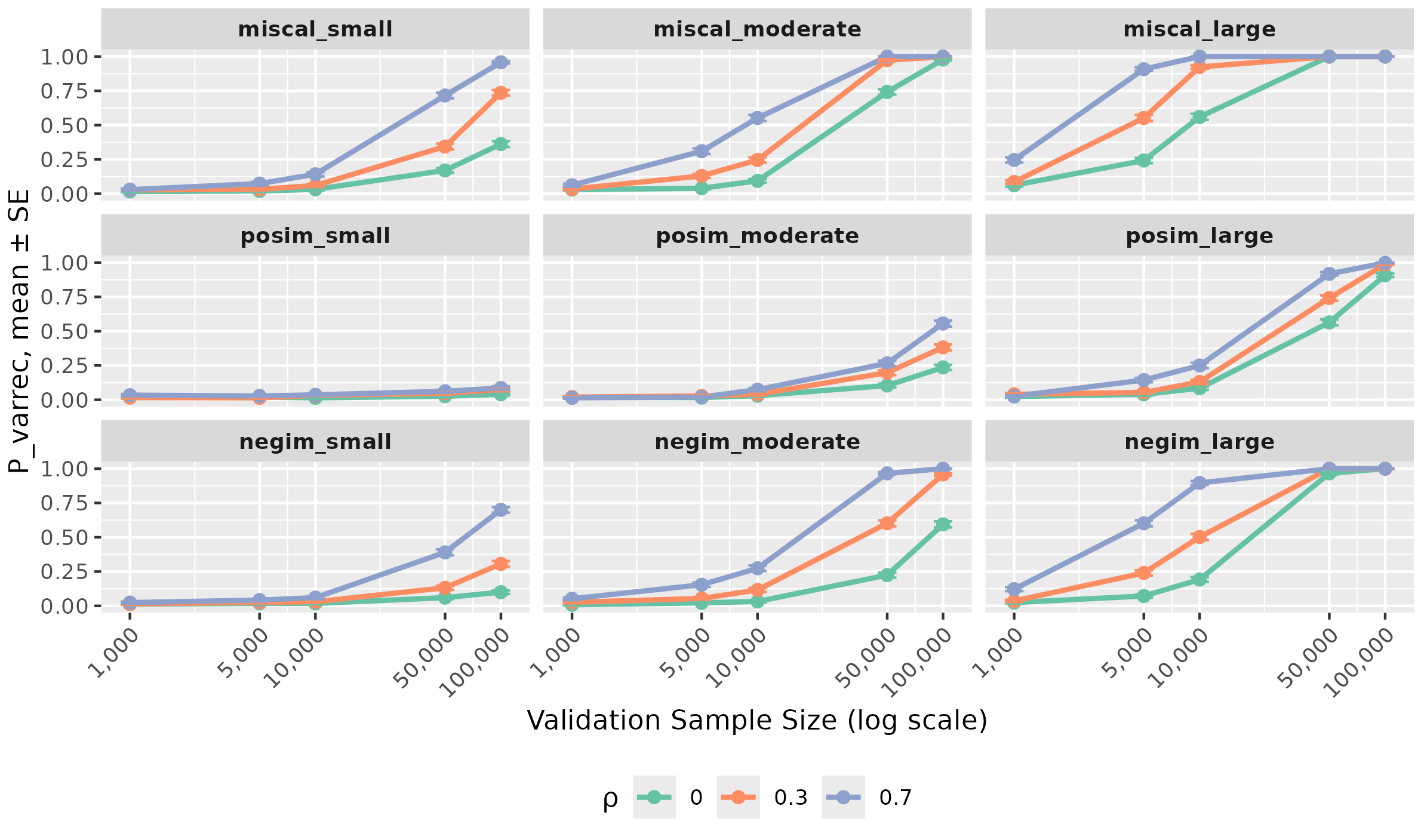}
\end{figure}

\begin{figure}[!h]
\centering
\caption{
Probability of recovering the true interaction driving unfairness under non-null settings as a function of sample size. Panels correspond to combinations of unfairness mechanism and severity. Lines represent Monte Carlo averages across $B=500$ replicates. Interaction recovery increases with both sample size and severity.
}
\label{fig:nonnull_interaction_recovery}

\includegraphics[width=\textwidth]{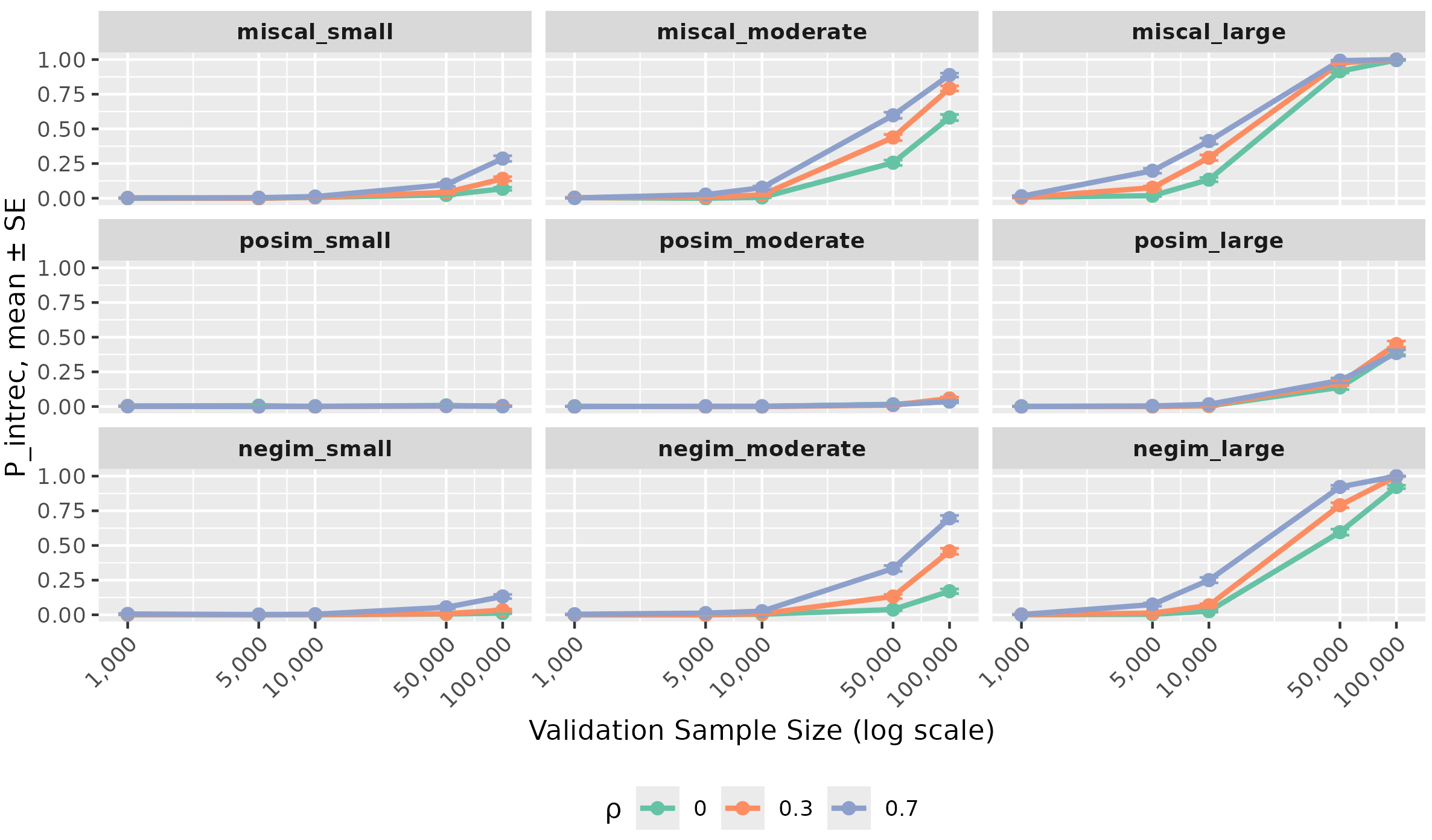}
\end{figure}

\begin{figure}[!h]
\centering
\caption{
Bias of leaf-wise calibration discrepancy estimates as a function of sample size. Panels correspond to combinations of unfairness mechanism (miscalibration, negative imbalance, positive imbalance) and severity (none, small, moderate, large). Lines show Monte Carlo averages over $B=500$ replicates, with error bars indicating Monte Carlo uncertainty. Bias is small across settings and decreases toward zero with increasing sample size, suggesting that the utree provides approximately unbiased estimates of subgroup-level calibration discrepancy.
}
\label{fig:nonnull_bias}

\includegraphics[width=\textwidth]{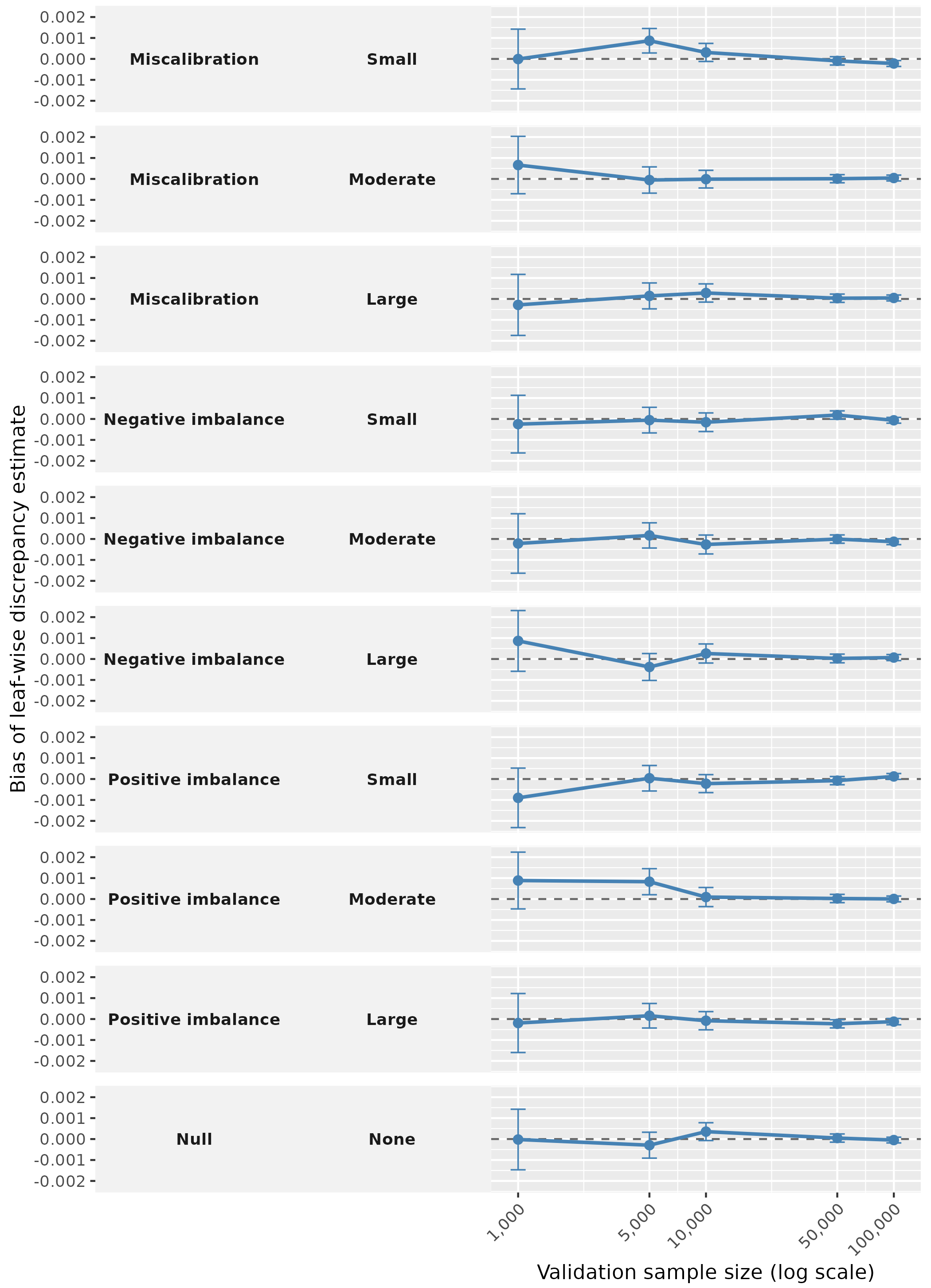}
\end{figure}

\begin{figure}[!h]
\centering
\caption{
Computation time (minutes) for fitting the utree under non-null settings as a function of sample size. Panels correspond to combinations of unfairness mechanism and severity. Lines represent Monte Carlo averages across $B=500$ replicates. Computation time increased approximately linearly over the range of sample sizes considered, suggesting the utree remains computationally feasible even at large sample sizes.
}
\label{fig:nonnull_fit_time}

\includegraphics[width=\textwidth]{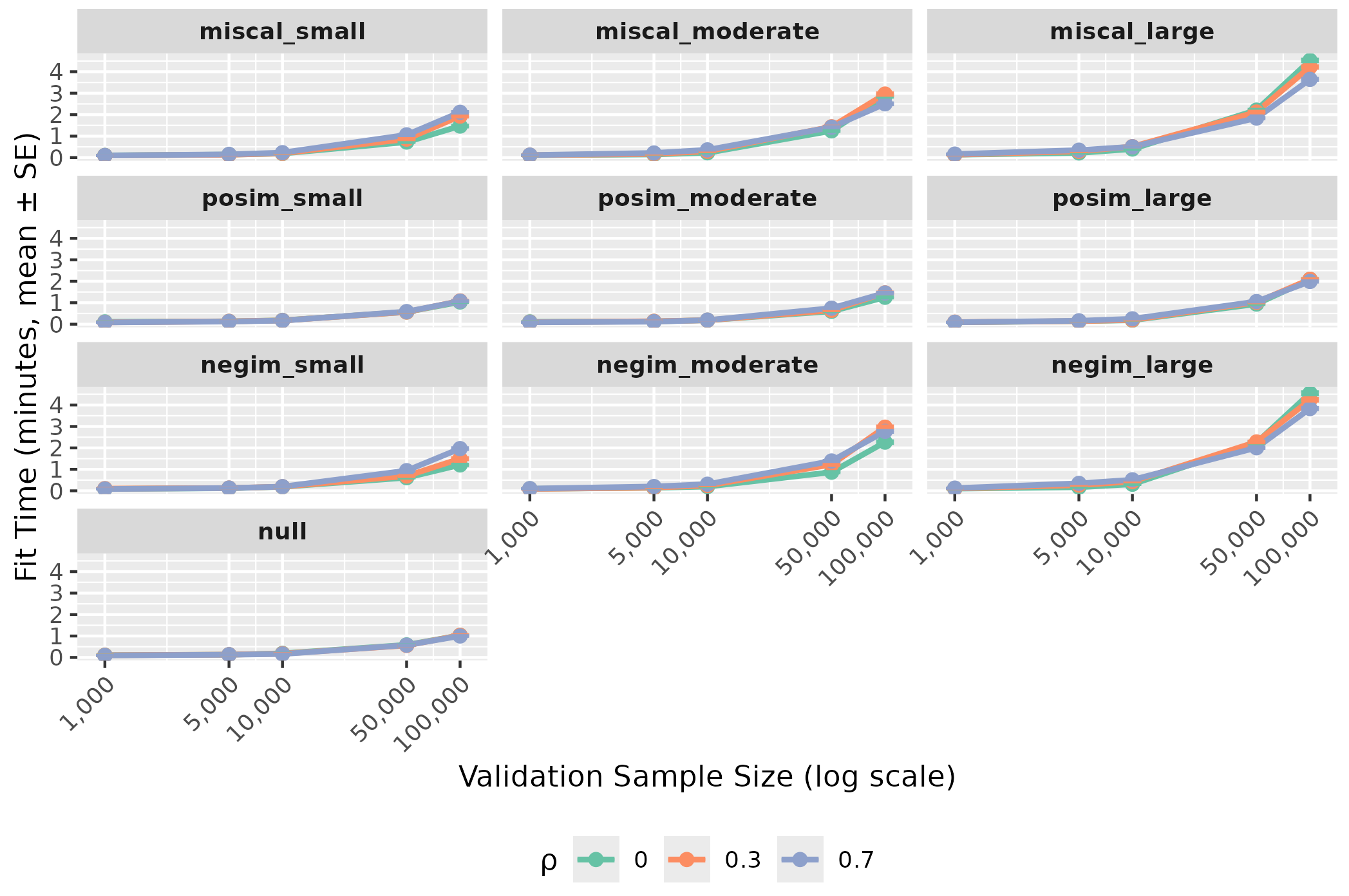}
\end{figure}

Additional simulation results characterize the data generation process \ref{pre_specified_approach} and the statistical behavior of the utree. Analyses evaluate recovery of unfairness-generating variables (Figure \ref{fig:nonnull_var_recovery}) and interactions (Figure \ref{fig:nonnull_interaction_recovery}), bias in subgroup discrepancy estimation (Figure \ref{fig:nonnull_bias}), with precise values reported in \ref{tab:cf_obs_comparison}. Computational scaling behavior is also assessed (Figure \ref{fig:nonnull_fit_time}).

\begin{table}[!h]
\centering
\caption{Summary of unfairness-generating mechanisms across simulation settings. For each mechanism (miscalibration, positive imbalance, negative imbalance) and severity level (none, small, moderate, large), results summarize $B=500$ Monte Carlo replicates of evaluation sets with $N=10{,}000$ and correlation $\rho=0.3$. Entries report the Monte Carlo mean (standard deviation) of subgroup-specific AUC ratios and calibration intercept differences relative to the reference subgroup $(Z_9,Z_{10})=(0,0)$. The AUC ratio is $\text{AUC}_g / \text{AUC}_{(0,0)}$, so values greater than 1 indicate improved discrimination relative to the reference subgroup. The calibration intercept difference is $\hat{\alpha}_g - \hat{\alpha}_{(0,0)}$, where $\hat{\alpha}$ is obtained from a logistic calibration model of $Y_i^0$ on $\mathrm{logit}(\tilde{s}_i)$; positive values indicate greater overestimation relative to the reference subgroup. AUC ratios remain near 1 under the miscalibration mechanism because rank ordering is largely preserved. In contrast, imbalance mechanisms modify predicted risks conditionally on the outcome, altering rank ordering and therefore discrimination.}
\label{pre_specified_approach}

\textbf{AUC ratio}\\[0.3em]

\begin{tabular}{lccc}
\toprule
Scenario & (1,0) & (0,1) & (1,1) \\
\midrule
null            & 0.999 (0.018) & 0.999 (0.018) & 1.002 (0.017)\\
miscal\_small   & 1.000 (0.017) & 1.000 (0.017) & 1.004 (0.016)\\
miscal\_moder   & 0.999 (0.017) & 0.999 (0.017) & 1.003 (0.016)\\
miscal\_large   & 0.999 (0.018) & 1.000 (0.018) & 1.002 (0.015)\\
posim\_small    & 1.012 (0.017) & 1.007 (0.017) & 1.035 (0.016)\\
posim\_moder    & 1.026 (0.017) & 1.013 (0.017) & 1.066 (0.015)\\
posim\_large    & 1.051 (0.017) & 1.025 (0.017) & 1.117 (0.015)\\
negim\_small    & 0.985 (0.018) & 0.993 (0.017) & 0.968 (0.017)\\
negim\_moder    & 0.971 (0.017) & 0.987 (0.018) & 0.931 (0.017)\\
negim\_large    & 0.943 (0.018) & 0.972 (0.018) & 0.855 (0.018)\\
\bottomrule
\end{tabular}

\vspace{1em}

\textbf{Calibration difference}\\[0.3em]

\begin{tabular}{lccc}
\toprule
Scenario & (1,0) & (0,1) & (1,1) \\
\midrule
null            & 0.004 (0.066) & -0.001 (0.070) & 0.001 (0.063)\\
miscal\_small   & -0.050 (0.068) & -0.022 (0.067) & -0.123 (0.065)\\
miscal\_moder   & -0.102 (0.069) & -0.053 (0.065) & -0.249 (0.064)\\
miscal\_large   & -0.200 (0.068) & -0.095 (0.068) & -0.500 (0.063)\\
posim\_small    & -0.014 (0.071) & 0.002 (0.070) & 0.012 (0.069)\\
posim\_moder    & -0.018 (0.068) & -0.010 (0.070) & 0.010 (0.070)\\
posim\_large    & -0.047 (0.070) & -0.019 (0.065) & -0.036 (0.070)\\
negim\_small    & -0.044 (0.066) & -0.025 (0.067) & -0.140 (0.061)\\
negim\_moder    & -0.079 (0.064) & -0.043 (0.067) & -0.264 (0.064)\\
negim\_large    & -0.154 (0.069) & -0.080 (0.068) & -0.464 (0.063)\\
\bottomrule
\end{tabular}
\end{table}

\begin{table}[ht]
\centering
\caption{Simulation results for the counterfactual unfairness tree (utree) under exchangeable correlation $\rho = 0.3$ and sample size $N = 10^3$. Rows correspond to the null scenario and the mechanism--severity unfairness scenarios (miscalibration, positive imbalance, and negative imbalance at small, moderate, and large severity). $P(\mathrm{split})$ is the proportion of replications in which the tree made at least one split, $P(\mathrm{var})$ is the proportion recovering at least one true unfairness-driving variable, $P(\mathrm{int})$ is the proportion recovering the true interaction, and Bias summarizes test set discrepancy prediction performance. Entries are reported as mean (SE) across replications.} 
\begin{tabular}{lcccc}

 & \textbf{P\_split} & \textbf{P\_var} & \textbf{P\_int} & \textbf{Bias} \\ 
  \hline
    null & 0.048 (0.010) & 0.000 (0.000) & 0.000 (0.000) & 0.000 (0.000) \\ 
miscal\_small & 0.112 (0.014) & 0.060 (0.011) & 0.008 (0.004) & 0.001 (0.000) \\ 
  miscal\_moderate & 0.358 (0.021) & 0.246 (0.019) & 0.024 (0.007) & -0.001 (0.000) \\ 
  miscal\_large & 0.940 (0.011) & 0.924 (0.012) & 0.292 (0.020) & -0.000 (0.000) \\ 
  posim\_small & 0.056 (0.010) & 0.032 (0.008) & 0.000 (0.000) & -0.000 (0.000) \\ 
  posim\_moderate & 0.086 (0.013) & 0.040 (0.009) & 0.000 (0.000) & 0.001 (0.000) \\ 
  posim\_large & 0.166 (0.017) & 0.130 (0.015) & 0.006 (0.003) & -0.000 (0.000) \\ 
  negim\_small & 0.086 (0.013) & 0.030 (0.008) & 0.002 (0.002) & -0.000 (0.000) \\ 
  negim\_moderate & 0.206 (0.018) & 0.116 (0.014) & 0.008 (0.004) & -0.000 (0.000) \\ 
  negim\_large & 0.626 (0.022) & 0.502 (0.022) & 0.068 (0.011) & 0.000 (0.000) \\ 

   \hline
\end{tabular}

\label{tab:cf_obs_comparison}
\end{table}

\FloatBarrier
\section{GUSTO-I Data Analysis Supplement}

\begin{figure}[htbp]
    \centering
    \includegraphics[width=0.85\textwidth]{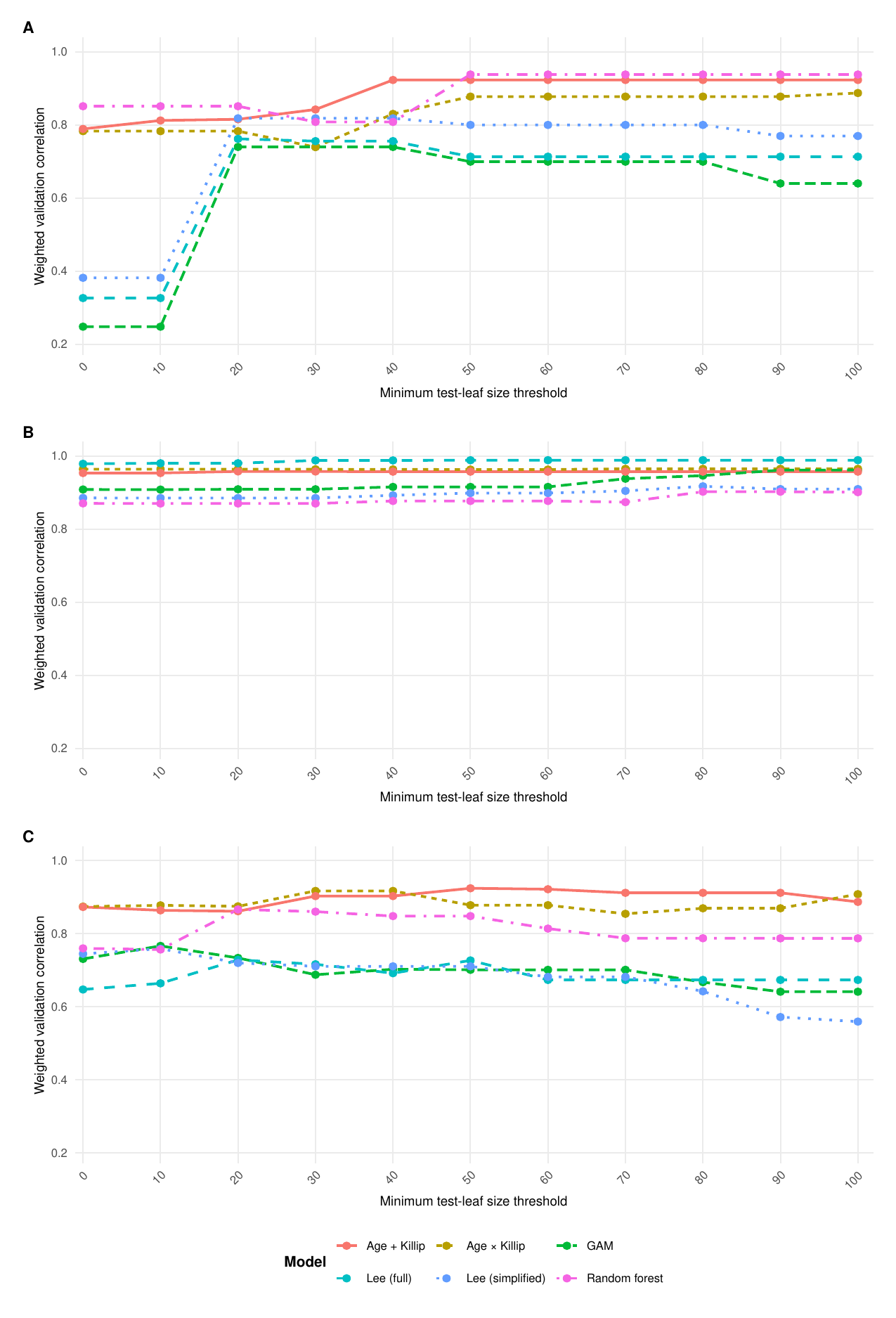}
    \caption{
    Sensitivity of out-of-sample validation to minimum terminal leaf-size thresholds for (A) calibration discrepancy, (B) true-positive rate (TPR), and (C) precision. Weighted correlations between evaluation-set and independently re-estimated test-set terminal-leaf performance are shown as increasingly small terminal leaves are excluded from validation. Higher correlations indicate stronger agreement in subgroup-specific performance across independent data splits. Validation was generally robust across a range of minimum leaf-size thresholds, although sensitivity varied by performance metric and prediction model. A minimum terminal leaf size of $n \geq 50$ provided a practical balance between validation stability and subgroup retention, particularly for calibration-based unfairness trees.
    }
    \label{fig:leaf_size_sensitivity}
\end{figure}

Table~\ref{tab:gusto_model_performance} summarizes overall discrimination and calibration performance for each mortality prediction model evaluated in the study. Although aggregate predictive performance was broadly similar across models, the utree analyses revealed meaningful heterogeneity in subgroup-level discrepancy structure beyond what was apparent from global metrics alone. Table~\ref{tab:utree_variable_dictionary} defines abbreviated variable labels used in subgroup visualizations and tree summaries. Figure~\ref{fig:leaf_size_sensitivity} evaluates the stability of subgroup discrepancy estimates as small leaves were excluded.

% latex table generated in R 4.3.0 by xtable 1.8-4 package
% Sat Apr 25 10:32:42 2026
\begin{table}[ht]
\centering
\caption{
Model performance in the GUSTO evaluation set. Discrimination is summarized by the area under the receiver operating characteristic curve (AUC), and calibration is summarized using the logistic recalibration model $\mathrm{logit}\{P(Y=1)\} = \alpha + \beta\,\mathrm{logit}(\hat s)$. Values in parentheses are 95\% confidence intervals (DeLong for AUC; Wald intervals for calibration parameters). An intercept of 0 and slope of 1 indicate ideal calibration. Intercepts greater than 0 indicate observed risk exceeds predicted risk on average (underprediction). Slopes less than 1 indicate overly extreme predictions, generally consistent with overfitting. Discrimination was similar across models (AUC approximately 0.78--0.82), with generally good calibration overall.
}
\label{tab:gusto_model_performance}
\begin{tabular}{lccc}
  \hline
  \\[-1.8ex]
 & \multicolumn{1}{c}{Discrimination} & \multicolumn{2}{c}{Calibration} \\
\cline{2-2} \cline{3-4}
Model & AUC (95\% CI) & Intercept (95\% CI) & Slope (95\% CI) \\
\hline
Logit: Age+Killip & 0.778 (0.764, 0.791) & 0.179 (0.029, 0.330) & 1.086 (1.023, 1.150) \\ 
  Logit: Age*Killip & 0.780 (0.766, 0.793) & 0.149 (0.001, 0.298) & 1.075 (1.012, 1.139) \\ 
  Logit: Lee simplified & 0.813 (0.801, 0.825) & -0.085 (-0.206, 0.036) & 0.956 (0.907, 1.006) \\ 
  Logit: Lee full & 0.813 (0.801, 0.825) & -0.093 (-0.213, 0.028) & 0.953 (0.904, 1.003) \\ 
  GAM & 0.815 (0.803, 0.827) & -0.097 (-0.218, 0.023) & 0.951 (0.903, 1.000) \\ 
  Random forest & 0.798 (0.785, 0.811) & -0.232 (-0.355, -0.111) & 0.927 (0.873, 0.981) \\ 
   \hline
\end{tabular}
\end{table}

\begin{table}[htbp]
\centering
\caption{Abbreviated labels used for candidate split variables in GUSTO-I analysis.}
\label{tab:utree_variable_dictionary}
\begin{tabular}{ll}
\toprule
Short name & Definition \\
\midrule
USe   & U.S. enrollment indicator \\
TTR   & Time to relief of chest pain $>$ 1 hour \\
ANT   & Anterior infarct location indicator \\
OTH   & Other infarct location indicator \\
SBP   & Systolic blood pressure \\
PULSE & Heart rate (pulse) \\
PMI   & Prior myocardial infarction \\
CVD   & Prior cardiovascular disease \\
CABG  & Prior coronary artery bypass graft \\
AGE   & Age (years) \\
F     & Female sex indicator \\
HT    & Height \\
WT    & Weight \\
HTN   & Hypertension history \\
K1    & Killip class I indicator \\
K3    & Killip class III indicator \\
K4    & Killip class IV indicator \\
KillipII & Killip class II indicator \\
smkquit    & Former smoker indicator \\
smkcurrent & Current smoker indicator \\
\bottomrule
\end{tabular}
\end{table}

\begin{figure}[htbp]
    \centering
    \includegraphics[width=\textwidth]{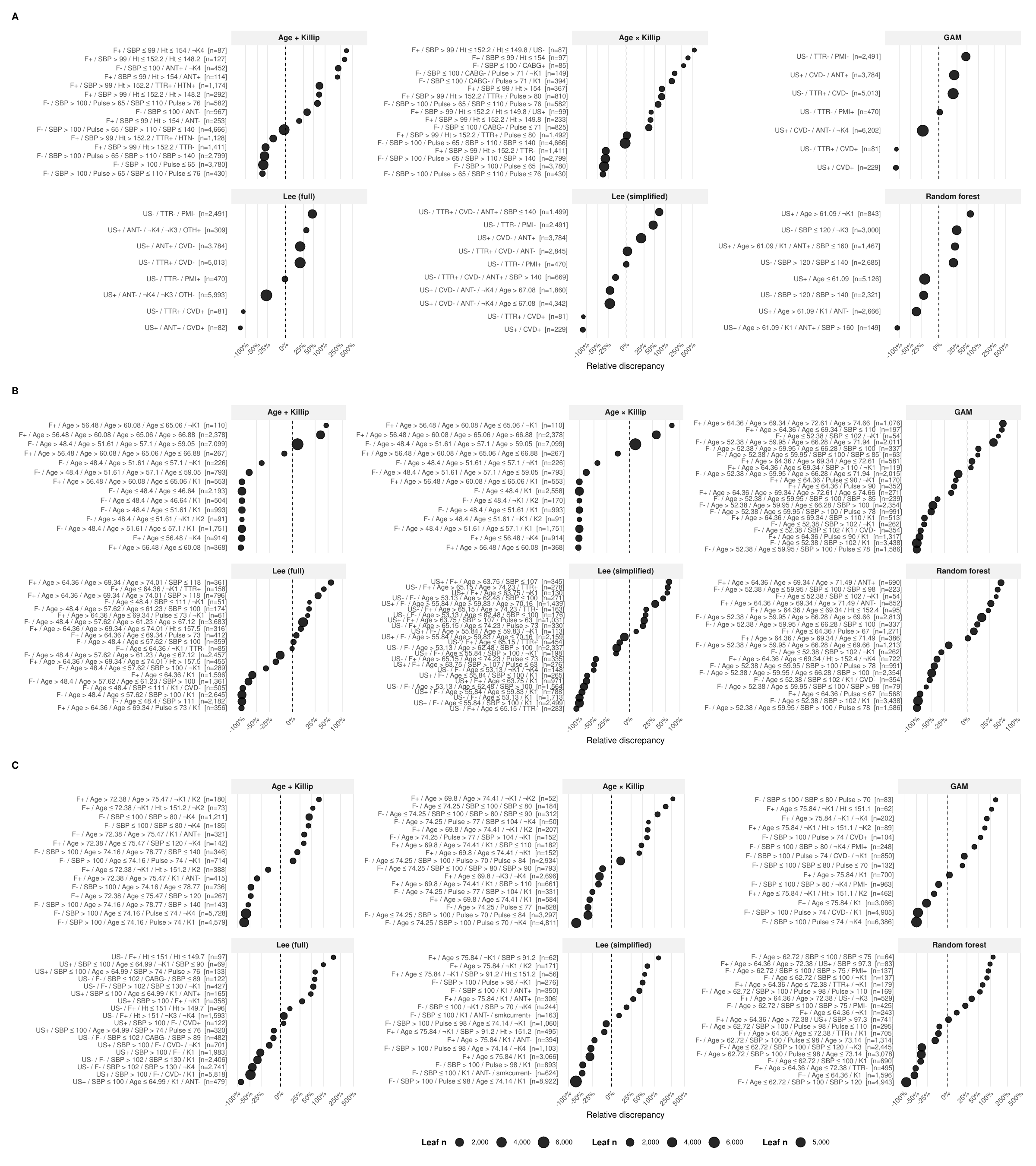}
    \caption{Terminal subgroups identified by unfairness trees fit to each prediction model using (A) calibration discrepancy, (B) true-positive rate (TPR), and (C) precision as the splitting criterion. Points denote terminal subgroups positioned by relative subgroup performance. For calibration, the horizontal axis shows the relative discrepancy $100\times\mathbb{E}(\hat{Y}_0-\hat{s})/\mathbb{E}(\hat{s})$, where $\mathbb{E}(\hat{s})$ is the model-wide average predicted risk. Positive values indicate underprediction ($\hat{Y}_0>\hat{s}$). For TPR and precision, values represent percent deviations from the corresponding model-wide performance metric, with positive (negative) values indicating better (worse) subgroup performance than the model overall. Point size is proportional to subgroup size. Terminal nodes with fewer than 50 observations are omitted. Dashed vertical lines denote zero discrepancy (or average model performance for TPR and precision). Subgroup signatures summarize the sequence of splits defining each terminal node; ``+'' and ``$-$'' denote the presence or absence of binary characteristics.}
    \label{fig:utree_leaf_discrepancy_all_metrics}
\end{figure}

\newpage

\end{document}